\documentclass[smallextended]{article}
\usepackage[english]{babel}
\usepackage[dvips]{graphicx}
\usepackage{amsthm}
\usepackage{exscale}
\usepackage{amsfonts}
\usepackage{amssymb}
\usepackage[intlimits,sumlimits]{amsmath}
\usepackage{amsmath,cases}
\usepackage[latin1]{inputenc}
\usepackage{mathrsfs}
\usepackage{fancyhdr}
\usepackage[misc,geometry]{ifsym}

\theoremstyle{definition}

\newtheorem{thm}{Theorem}

\newtheorem{lm}[thm]{Lemma}

\newtheorem{defin}[thm]{Definition}

\newcommand{\ab}{\mathbf{a}}
\newcommand{\bb}{\mathbf{b}}
\newcommand{\xb}{\mathbf{x}}

\newcommand{\vb}{\mathbf{v}}
\newcommand{\wb}{\mathbf{w}}
\newcommand{\ub}{\mathbf{u}}
\newcommand{\zb}{\mathbf{z}}
\newcommand{\xib}{\boldsymbol{\xi}}
\newcommand{\psib}{\boldsymbol{\psi}}

\newcommand{\omb}{\boldsymbol{\omega}}
\newcommand{\ud}{\mathrm{d}}

\newcommand{\unifSo}{u_{S^2}(\ud \boldsymbol{\omega})}

\newcommand{\intethree}{\int_{\mathbb{R}^3}}
\newcommand{\lnorm}{\mid\mid \!}
\newcommand{\rnorm}{\! \mid\mid}

\newcommand{\ind}{1 \! \textrm{l}}

\newcommand{\borelthree}{\mathscr{B}(\mathbb{R}^3)}
\newcommand{\rone}{\mathbb{R}}
\newcommand{\rthree}{\mathbb{R}^3}

\newcommand{\Cb}{\mathrm{C}_{b}(\mathbb{R}^3)}

\newcommand{\treen}{\mathfrak{t}_n}
\newcommand{\treenk}{\mathfrak{t}_{n, k}}

\newcommand{\tree}{\mathfrak{t}}

\newcommand{\DD}{\mathrm{C}^{2}_{b}(\mathbb{R}^3)}
\newcommand{\mut}{\mu(\cdot, t)}

\newcommand{\pt}{\textsf{P}_t}
\newcommand{\et}{\textsf{E}_t}

\author{Emanuele Dolera \footnote{Address: Dipartimento di Scienze Fisiche, Informatiche e Matematiche, via Campi 213/b, 41125 Modena, Italy. E-mail: emanuele.dolera@unimore.it} \\
\emph{Universit\`a di Modena e Reggio Emilia}}

\title{\textbf{Mathematical treatment of the homogeneous Boltzmann equation for Maxwellian molecules in the presence of singular kernels}}
\date{}

\begin{document}
\maketitle

\begin{abstract}
This paper proves the existence of weak solutions to the the spatially homogeneous Boltzmann equation for Maxwellian molecules, when the initial data are chosen from the space of all Borel probability measures on $\rthree$ with finite second moments and the (angular) collision kernel satisfies a very weak cutoff condition, namely $\int_{-1}^{1} x^2 b(x) \ud x < +\infty$. Conservation of momentum and energy is also proved for these weak solutions, without resorting to any boundedness of the entropy.
\end{abstract}

\small{\textbf{Keywords}: Boltzmann equation; Maxwellian molecules; moments; sum of random variables; uniform integrability; very weak cutoff; weak solution} \\

\section{Introduction and main results} \label{sect:intro}

This paper deals with the spatially homogeneous Boltzmann equation for \emph{Maxwellian molecules} (SHBEMM), commonly written as
\begin{eqnarray}
\frac{\partial}{\partial t} f(\vb, t) &=& \intethree\int_{S^2} [f(\vb_{\ast}, t)
f(\wb_{\ast}, t) \ - \ f(\vb, t) f(\wb, t)] \times \nonumber \\
&\times& b\left(\frac{\wb - \vb}{|\wb - \vb|} \cdot \omb \right) \unifSo \ud \wb \label{eq:boltzmann}
\end{eqnarray}
with $(\vb, t) \in \rthree \times (0, +\infty)$. Existence and evolution of low-order moments of its solutions
are the main topics at issue, in the event that \emph{grazing collisions} are significantly taken into account. In spite of a vast literature on the subject, very few papers aim at minimizing as much as possible the set of hypotheses on both the initial datum and the collision kernel within a rigorous mathematical framework, as this work intends to do.

As to the symbols in (\ref{eq:boltzmann}), $u_{S^2}$ denotes the uniform measure (i.e. the normalized Riemannian measure) on the unit sphere $S^2$, embedded in $\rthree$. The post-collisional velocities $\vb_{\ast}$ and $\wb_{\ast}$ are defined according to the $\omb$-representation:
\begin{equation} \label{eq:omrepresentation}
\begin{array}{lll}
\vb_{\ast} &:= \ \vb &+ \ \ [(\wb - \vb) \cdot \omb] \ \omb  \\
\wb_{\ast} &:= \ \wb &- \ \ [(\wb - \vb) \cdot \omb] \ \omb
\end{array}
\end{equation}
where $\cdot$ designates the standard scalar product. The solution $f(\vb, t)$ is a probability density function (w.r.t. the $\vb$-variable, for every $t$) which characterizes the probability law of a single molecule's velocity, randomly chosen in a chaotic bath of like molecules. See \cite{cerS,cip,vil} for an exhaustive explanation. The \emph{(angular) collision kernel} $b$ is an even measurable function from $[-1, 1]$ into $[0, +\infty]$ which plays a central role in the study of the Maxwellian molecules. Originally, this name was reserved for molecules repelling each other with a force inversely proportional to the fifth power of their distance, after Maxwell had evaluated the exact expression of $b$ in this peculiar case. See \cite{max} and Section 3 of Chapter 2A of \cite{vil}. Nowadays, the word Maxwellian indicates the presence of a kernel depending only on $\frac{\wb - \vb}{|\wb - \vb|} \cdot \omb$, as in (\ref{eq:boltzmann}), and the aim is to investigate in which way such a function may influence the relative solution of the SHBEMM. The present work deals with collision kernels satisfying
\begin{equation} \label{eq:vwcutoff}
\int_{-1}^{1} x^2 b(x) \ud x < +\infty
\end{equation}
i.e. a \emph{very weak angular cutoff}, which is the weakest assumption on $b$ considered so far in the literature, starting from \cite{desvKac}. Condition (\ref{eq:vwcutoff}) is fulfilled when $b$ assumes the particular expression found out by Maxwell, which possesses a unique singularity at $x = 0$ in such a way that $b(x) \sim |x|^{-3/2}$, just revealing the presence of grazing collisions. The last important object is the \emph{initial datum}, to be considered throughout this paper as an element of the class of all Borel probability measures (p.m.'s) on $\rthree$ with finite second moments -- therefore not necessarily absolutely continuous and not constrained to any finite-entropy condition. In this framework, the first task consists in a weak reformulation of (\ref{eq:boltzmann}). The motivations to aim at such generality are both theoretical and practical: For example, in \cite{cl} it is expressly remarked that ``in view of statistical physics, initial data are best chosen from the largest class, say the positive, finite Borel measures on $\rthree$'', while in \cite{miwe} the authors underline the importance of dropping finite-entropy conditions ``since no control of entropy can be expected in the explicit Euler scheme''. In fact, a number of noteworthy papers, such as \cite{abra,cl,miwe}, succeeded in proving important facts without assuming the finiteness of the entropy of the initial datum.

The very weak cutoff condition, in conjunction with a minimization of the hypotheses on the initial datum, leads to study a larger class of solutions than the usual one, arising in the context of integrable or at least not too singular collision kernels. Actually, this enlargement makes the problems of existence and uniqueness more challenging from a mathematical point of view and
introduces new difficulties in determining the properties of such solutions. For example, a rigorous proof that these weak solutions preserve momentum and energy, in absence of extra-condition on the initial datum, is still lacking. This fact has even been doubted in \cite{desvKac}, where one wonders whether the energy may decrease. Besides, general initial data with the above-mentioned characteristics can be completely managed in the case that $b$ is summable (\emph{Grad cutoff assumption}), thanks to a consolidate knowledge on the subject which started with the works \cite{bob88,iktr,mck7,mor,wil} and culminated with \cite{puto,tovil}. The same extension in the \emph{weak cutoff} case, which corresponds to assuming $\int_{-1}^{1} |x| b(x) \ud x < +\infty$, is treated in \cite{caka,ta,tovil}. Coming to the case of kernels satisfying (\ref{eq:vwcutoff}), a general line of reasoning to tackle existence questions was devised by Arkeryd \cite{ark}, who considered a sequence of integrable truncations of the kernel $b$, say $\{b_n\}_{n\geq 1}$, to obtain a sequence of auxiliary solutions approximating the real (unknown) solution. One of the main difficulties in the Arkeryd approach is to show some weak compactness of the approximating sequence, in order to get a converging subsequence. Actually, the more natural form of compactness in Boltzmann's equation can be derived from the boundedness of the entropy, as successfully done in pioneering works such as \cite{desvKac,fou,goudon,villNew}. On the other hand, when the initial datum is a p.m. with finite second moments, not constrained to a finite-entropy condition, the only available form of compactness ought to be derived  from the conservation of momentum and energy, as first proposed in \cite{puto} and then developed in \cite{caka}. In the wake of this line of research, the present work proposes a weak reformulation of (\ref{eq:boltzmann}) which fits the Arkeryd approach, with the contrivance to corroborate the weak compactness with a form of uniform integrability of the second absolute moments of the approximating solutions. This last achievement answers an open question in \cite{cl} about the validity of such uniform integrability under minimal assumptions on the collision kernel. The new proof given here is based on a probabilistic representation of the solutions -- recently proposed in \cite{doreACTA} and summed up in Subsection \ref{sect:grad} of this paper -- which turns out to be particulary effective for generalizing inequalities about the uniform boundedness (w.r.t. time) of the moments of the solutions. As a last remark about the placement of these results in the literature, it is right to comment on the recent paper \cite{morimoto}, appeared when the present article was a first draft, as the statements of existence and uniqueness contained therein are rather similar to those in Theorem \ref{thm:main} below, even if \cite{morimoto} starts from a different weak formulation. Indeed, my original aim was twofold: To complete some points expressly mentioned in \cite{doreACTA}, and to seize this opportunity to deal with those points within a framework more general than the required one. I have decided to carry through my own work even after the publication of \cite{morimoto} since I found the short proof therein not completely satisfactory. In fact, it seems not clear what kind of integral the author is adopting: On the one hand, Lemma 2.2 turns out to be false if integrals are of Lebesgue type (see the remark after Lemma \ref{lm:morimoto} of the present paper). On the other hand, if they are thought of as improper Riemann integrals, then there is a crucial -- i.e. the core of the proof -- exchange of limit with integral, immediately after (23), lacking in explanation. These shortcomings are not easy to restore, since this would require a kind of uniform convergence of the approximating sequence not yet proved. In any case, the problem can be solved, as I do here, by following a different strategy which shows, in addition, that the solutions (in the meaning of Definition \ref{defin:solution} below) conserve momentum and energy.

The rest of the paper is organized as follows: Subsection \ref{sect:weak} presents a distinguished weak form of the SHBEMM; Subsection \ref{sect:grad} gathers some well-known facts about the SHBEMM, which are valid under the Grad cutoff hypothesis, and states a new proposition about the uniform integrability of the second absolute moments of the solutions; Subsection \ref{sect:main} formulates the main results of existence and evolution of the moments of the solutions of the SHBEMM with a singular kernel; finally, Section \ref{sect:proofs} contains the proofs.

\subsection{Weak form of the equation} \label{sect:weak}

The first point of the program concerns the meaning of the RHS of (\ref{eq:boltzmann}) when $b$ satisfies (\ref{eq:vwcutoff}). In fact, an eventual singularity of the kernel could make the integrand a non-summable function even if $f(\vb, t)$ is very regular. See the illuminating discussion in Subsection 4.1 of Chapter 2B of \cite{vil}. The above-mentioned problem can be tackled by first considering the standard weak formulation
\begin{eqnarray}
\frac{\ud}{\ud t} \intethree \psi(\vb) \mu(\ud \vb, t) &=& \frac{1}{2} \intethree \intethree \int_{S^2} [\psi(\vb_{\ast}) + \psi(\wb_{\ast}) - \psi(\vb) - \psi(\wb)] \ind\{\vb \neq \wb\}
\times \nonumber \\
&\times& b\left(\frac{\wb - \vb}{|\wb - \vb|} \cdot \omb \right)
\unifSo \mu(\ud \vb, t) \mu(\ud \wb, t) \label{eq:wboltzmann}
\end{eqnarray}
which is derived from (\ref{eq:boltzmann}) by multiplying both sides by some regular function $\psi(\vb)$, integrating formally in the $\vb$-variable and putting $f(\vb, t)\ud \vb = \mu(\ud \vb, t)$. Now, the initial datum can be thought of as a Borel p.m. $\mu_0$ on $\rthree$, not necessarily absolutely continuous, and a solution of (\ref{eq:boltzmann}) is intended accordingly
as a family $\{\mut\}_{t \geq 0}$ of Borel p.m.'s on $\rthree$, such that $\mu(\cdot, 0) = \mu_0(\cdot)$ and satisfying (\ref{eq:wboltzmann}) in a certain sense. Equation (\ref{eq:wboltzmann}) would suffice to formulate a precise notion of solution in the case that $\int_{-1}^{1} |x|b(x) \ud x < +\infty $ holds, whilst, in the very weak cutoff context, some further steps are required, as already noted in \cite{goudon,villNew}. More precisely, when $\vb \neq \wb$, the integral
$$
\int_{S^2} [\psi(\vb_{\ast}) + \psi(\wb_{\ast}) - \psi(\vb) - \psi(\wb)] b\left(\frac{\wb - \vb}{|\wb - \vb|} \cdot \omb \right) \unifSo
$$
can be re-written by the (formal) change of variable $\omb \leftrightarrow (\theta, \varphi)$, given by
$$
\omb(\theta, \varphi, \ub) := \cos\theta \sin\varphi \ab(\ub) + \sin\theta \sin\varphi \bb(\ub) + \cos\varphi \ub
$$
where $(\theta, \varphi) \in (0, 2\pi) \times [0, \pi]$, $\ub := \frac{\wb - \vb}{|\wb - \vb|}$ and $\{\ab(\ub), \bb(\ub), \ub\}$ is an orthonormal basis of $\rthree$. The identities in (\ref{eq:omrepresentation}) become
\begin{eqnarray}
\vb_{\ast} &=& \vb + |\wb - \vb|\cos\varphi \omb(\theta, \varphi, \ub) \nonumber \\
\wb_{\ast} &=& \wb - |\wb - \vb|\cos\varphi \omb(\theta, \varphi, \ub) \nonumber
\end{eqnarray}
and so, after putting $x = \cos\varphi$, the Taylor formula with integral remainder yields
\begin{eqnarray}
\psi(\vb_{\ast}(x)) &=& \psi(\vb) + x \left(\nabla\psi(\vb) \cdot \frac{\ud \vb_{\ast}}{\ud x}(0)\right) + x^2\int_{0}^{1} (1 - s) \frac{\ud^2 \psi(\vb_{\ast})}{\ud x^2}(sx) \ud s \nonumber \\
\psi(\wb_{\ast}(x)) &=& \psi(\wb) + x \left(\nabla\psi(\wb) \cdot \frac{\ud \wb_{\ast}}{\ud x}(0)\right) + x^2\int_{0}^{1} (1 - s) \frac{\ud^2 \psi(\wb_{\ast})}{\ud x^2}(sx) \ud s \ . \nonumber
\end{eqnarray}
Concerning the first-order terms, observe that $\frac{\ud \vb_{\ast}}{\ud x}(0)$ and $\frac{\ud \wb_{\ast}}{\ud x}(0)$ are given, up to a factor $\pm |\wb - \vb|$, by $\cos\theta \ab(\ub) + \sin\theta \bb(\ub)$, so that
$$
\int_{0}^{2\pi} \left(\nabla\psi(\vb) \cdot \frac{\ud \vb_{\ast}}{\ud x}(0)\right) \ud\theta = \int_{0}^{2\pi} \left(\nabla\psi(\wb) \cdot \frac{\ud \wb_{\ast}}{\ud x}(0)\right) \ud\theta = 0\ .
$$
The second-order terms can be treated by means of the following
\begin{lm} \label{lm:villani}
\emph{Let} $b$ \emph{satisfy} (\ref{eq:vwcutoff}) \emph{and let} $\chi$ \emph{any Borel p.m. on} $\rthree$ \emph{such that} $\intethree |\vb|^2 \chi(\ud \vb) < +\infty$. \emph{Then},
\begin{eqnarray}
\intethree \intethree \int_{-1}^{1}\int_{0}^{2\pi} \int_{0}^{1}&& \left[
\Big{|}\frac{\ud \vb_{\ast}}{\ud x}(s \xi)\Big{|}^2 + \Big{|}\frac{\ud \wb_{\ast}}{\ud x}(s \xi)\Big{|}^2 + \Big{|}\frac{\ud^2 \vb_{\ast}}{\ud x^2}(s \xi)\Big{|} + \Big{|}\frac{\ud^2 \wb_{\ast}}{\ud x^2}(s \xi)\Big{|}\right] \times \nonumber \\
&\times&  \ind\{\vb \neq \wb\} (1-s) \xi^2 b(\xi) \ud s\ud\theta\ud\xi \chi(\ud \vb) \chi(\ud \wb) < +\infty \ . \nonumber
\end{eqnarray}
\end{lm}
Hence, given a twice differentiable function $\psi$ with bounded derivatives up to the order two (henceforth indicated by $\psi \in \DD$), the precise mathematical meaning of the RHS in (\ref{eq:wboltzmann}) is
\begin{eqnarray}
&& \frac{1}{8\pi} \intethree \intethree \int_{-1}^{1}\int_{0}^{2\pi} \int_{0}^{1}  \ind\{\vb \neq \wb\} \xi^2 (1 - s) \Big{[} \nabla \psi(\vb_{\ast}(s \xi)) \cdot \frac{\ud^2 \vb_{\ast}}{\ud x^2}(s \xi) \nonumber \\
&+& \nabla \psi(\wb_{\ast}(s \xi)) \cdot \frac{\ud^2 \wb_{\ast}}{\ud x^2}(s \xi) + \left(\frac{\ud \vb_{\ast}}{\ud x}(s \xi)\right)^t \mathrm{Hess}[\psi](\vb_{\ast}(s \xi))\left(\frac{\ud \vb_{\ast}}{\ud x}(s \xi)\right)  \nonumber \\
&+& \left(\frac{\ud \wb_{\ast}}{\ud x}(s \xi)\right)^t \mathrm{Hess}[\psi](\wb_{\ast}(s \xi))\left(\frac{\ud \wb_{\ast}}{\ud x}(s \xi)\right)\Big{]} b(\xi) \ud s\ud\theta\ud\xi  \mu(\ud \vb, t) \mu(\ud \wb, t) \ \ \ \ \label{eq:vwRHS}
\end{eqnarray}
which constitutes the starting point for a rigorous definition.
\begin{defin}[Weak solution]\label{defin:solution}
\emph{Let} $b$ \emph{satisfy} (\ref{eq:vwcutoff}) \emph{and let} $\mu_0$ \emph{be a Borel p.m. on} $\rthree$ \emph{such that} $\intethree |\vb|^2 \mu_0(\ud \vb) < +\infty$. \emph{Then, a} weak solution of (\ref{eq:boltzmann}) \emph{is defined to be any family} $\{\mut\}_{t \geq 0}$ \emph{of Borel p.m.'s on} $\rthree$ \emph{such that}
\begin{enumerate}
\item[i)] $\mu(\cdot, 0) = \mu_0(\cdot)$;
\item[ii)] $t \mapsto \intethree \psi(\vb) \mu(\ud\vb, t)$ \emph{is continuous on} $[0,+\infty)$ \emph{and continuously differentiable on} $(0, +\infty)$, \emph{for all} $\psi \in \DD$;
\item[iii)] $\intethree |\vb|^2 \mu(\ud \vb, t) < +\infty$ \emph{for all} $t \geq 0$;
\item[iv)] $\mut$ \emph{satisfies} (\ref{eq:wboltzmann}) \emph{for all} $t > 0$ \emph{and for all} $\psi \in \DD$, \emph{provided that the RHS in} (\ref{eq:wboltzmann}) \emph{has the meaning specified by} (\ref{eq:vwRHS}).
\end{enumerate}
\end{defin}
It is worth noting that (\ref{eq:vwRHS}) has a precise mathematical meaning in view of point iii) and Lemma \ref{lm:villani}.
Furthermore, when $\int_{-1}^{1} |x| b(x) \ud x < +\infty$ holds, the RHS in (\ref{eq:wboltzmann}) is well-defined -- without invoking (\ref{eq:vwRHS}) -- for any test function $\psi$ which is bounded and Lipschitz continuous. As a consequence, in the weak cutoff context, any initial datum satisfying $\intethree |\vb| \mu_0(\ud \vb) < +\infty$ is allowed, with the proviso that condition iii) of the above definition is relaxed to $\intethree |\vb| \mu(\vb, t) < +\infty$ for all $t \geq 0$. See \cite{ta}. The Grad cutoff case is even simpler, as shown in the next subsection.

\subsection{The Grad cutoff case} \label{sect:grad}

When $b$ is a summable function, the treatment of (\ref{eq:boltzmann}) can be based on explicit formulas: A rare feature that has made the SHBEMM an attractive model to investigate, at least from a mathematical point of view. The problems of the existence, uniqueness and evolution of the moments were settled down long ago in \cite{iktr,mor,wil}, while an up-to-date revisiting is in \cite{caka,puto,tovil}. To start with, it is useful to sum up a number of well-known facts, with a view to their use for both stating and proving the new results. First, one can assume, without affecting the generality, the validity of the normalization
\begin{equation} \label{eq:cutoff}
\int_{S^2} b(\ub \cdot \omb) \unifSo = \int_{0}^{1} b(x) \ud x = 1\ \ \ \ \ (\forall\ \ub \in S^2)
\end{equation}
to re-write the SHBEMM as
$$
\frac{\partial}{\partial t} f(\vb, t) = Q[f(\cdot, t), f(\cdot, t)](\vb) - f(\vb)
$$
with $Q[p, q](\vb) := \intethree\int_{S^2} p(\vb_{\ast})q(\wb_{\ast}) b\left(\frac{\wb - \vb}{|\wb - \vb|} \cdot \omb \right) \unifSo \ud \wb$. Under (\ref{eq:cutoff}), the bilinear operator $Q$ sends the couple $(p, q)$ of probability densities into a new probability density on $\rthree$. Next, to include the case of initial data which are not absolutely continuous p.m.'s, define the operator $\mathcal{Q}$, which sends a pair $(\zeta, \eta)$ of Borel p.m.'s into a new Borel p.m. on $\rthree$, according to
\begin{equation} \label{eq:extendedQ}
\mathcal{Q}[\zeta, \eta](\ud \vb) := \text{w-lim}_{n \rightarrow \infty} Q[p_n, q_n](\vb) \ud \vb
\end{equation}
where $p_n$ ($q_n$, respectively) denotes the density of $\zeta_n$ ($\eta_n$, respectively), $\{\zeta_n\}_{n\geq 1}$ and $\{\eta_n\}_{n\geq 1}$ being two sequences of absolutely continuous p.m.'s such that $\zeta_n$ ($\eta_n$, respectively)
converges weakly to $\zeta$ ($\eta$, respectively). Recall that a statement as ``$\zeta_n$ converges weakly to $\zeta$''
($\zeta_n \Rightarrow \zeta$, in symbols) means that $\intethree \psi(\vb)\zeta_n(\ud \vb) \rightarrow \intethree \psi(\vb)\zeta(\ud \vb)$ for every bounded and continuous $\psi$ ($\psi \in \Cb$, in symbols). The following result states that $\mathcal{Q}[\zeta, \eta]$ is well-defined.
\begin{lm} \label{lm:extendedQ}
\emph{Let} $b$ \emph{meet} (\ref{eq:cutoff}). \emph{Then, the limit in} (\ref{eq:extendedQ}) \emph{exists and is independent of the choice of the approximating sequences} $\{\zeta_n\}_{n\geq 1}$ \emph{and} $\{\eta_n\}_{n\geq 1}$ \emph{and}
\begin{equation} \label{eq:bobylevQ}
\intethree \psi(\vb) \mathcal{Q}[\zeta, \eta](\ud\vb) = \intethree\intethree \int_{S^2} \psi(\vb_{\ast})
b\left(\frac{\wb - \vb}{|\wb - \vb|} \cdot \omb \right) \unifSo \zeta(\ud\vb) \eta(\ud\wb)
\end{equation}
\emph{holds for every} $\psi \in \Cb$. \emph{Moreover, if} $R$ \emph{is any orthogonal} $3 \times 3$ \emph{matrix and} $f_R$ \emph{denotes the linear map} $\vb \mapsto R\vb$, \emph{then}
\begin{equation} \label{eq:truccoR}
\mathcal{Q}[\zeta \circ f_{R}^{-1}, \eta \circ f_{R}^{-1}] = \mathcal{Q}[\zeta, \eta] \circ f_{R}^{-1}\ .
\end{equation}
\end{lm}
As a corollary, the following equation, known as \emph{Bobylev's identity} \cite{bob88},
\begin{equation} \label{eq:bobylev}
\hat{\mathcal{Q}}[\zeta, \eta](\xib) = \int_{S^2} \hat{\zeta}(\xib - (\xib \cdot \omb)\omb) \hat{\eta}((\xib \cdot \omb)\omb)  \ b\left(\frac{\xib}{|\xib|} \cdot \omb \right) \unifSo
\end{equation}
is valid for every $\xib \in \rthree\setminus\{\mathbf{0}\}$, where $\hat{}$ denotes the Fourier transform, according to
$\hat{\zeta}(\xib) := \intethree e^{i \xib \cdot \xb} \zeta(\ud\xb)$. With this notation at hand, one can put
\begin{equation} \label{eq:iterationQ}
\begin{array}{lll}
\mathcal{Q}_1[\mu_0] &:= \mu_0 & {} \\
\mathcal{Q}_n[\mu_0] &:= \frac{1}{n-1} \sum_{j=1}^{n-1} \mathcal{Q}\left[\mathcal{Q}_j[\mu_0], \mathcal{Q}_{n-j}[\mu_0]\right] & \ \ \ \ \text{for} \ n \geq 2\ ,
\end{array}
\end{equation}
to state the following
\begin{thm}\label{thm:cutoff}
\emph{Let} $b$ \emph{satisfy} (\ref{eq:cutoff}) \emph{and let} $\mu_0$ \emph{be any Borel p.m. on} $\rthree$. \emph{Then, the so-called} Wild sum
\begin{equation}\label{eq:wild}
\mu(\cdot, t) := \sum_{n = 1}^{+\infty} e^{-t} (1 - e^{-t})^{n-1} \mathcal{Q}_n[\mu_0](\cdot) \ \ \ \ \ (t \geq 0)
\end{equation}
\emph{is the only solution of} (\ref{eq:boltzmann}) \emph{that meets points} i)-ii)-iv) \emph{of Definition} \ref{defin:solution}
\emph{with initial datum} $\mu_0$, \emph{the test functions being chosen from} $\Cb$. \emph{Moreover, if} $R$ \emph{and} $f_R$ \emph{are as in Lemma} \ref{lm:extendedQ}, \emph{then} $\{\mu(\cdot, t) \circ f_{R}^{-1}\}_{t \geq 0}$ \emph{is the only solution of} (\ref{eq:boltzmann}) \emph{with} $\mu_0 \circ f_{R}^{-1}$ \emph{as initial datum}. \emph{Finally, if} $\mathfrak{m}_2 := \intethree |\vb|^2 \mu_0(\ud \vb) < +\infty$, \emph{then} $\{\mu(\cdot, t)\}_{t \geq 0}$ \emph{shares the following properties: First, momentum and kinetic energy are preserved, i.e.}
\begin{equation} \label{eq:Vu0}
\intethree \vb \mu(\ud \vb, t) = \intethree \vb \mu_0(\ud \vb) := \overline{\mathbf{V}} \ \ \ and \ \ \ \intethree |\vb|^2 \mu(\ud \vb, t) = \mathfrak{m}_2
\end{equation}
\emph{are in force for all} $t \geq 0$. \emph{Second},
\begin{equation} \label{eq:ViVj}
\lim_{t \rightarrow +\infty} \Big{[}\intethree v_i v_j \mu(\ud \vb, t) - \overline{V}_i \overline{V}_j\Big{]} = \frac{\delta_{ij}}{3}\intethree |\vb - \overline{\mathbf{V}}|^2 \mu_0(\ud\vb)
\end{equation}
\emph{holds for every} $i, j \in\{1, 2, 3\}$, $\delta_{ij}$ \emph{standing for the Kronecker delta. Third, there exist a positive constant} $C(\mu_0)$ \emph{and a continuous, non-decreasing function} $q : [0,+\infty) \rightarrow [0,+\infty)$, \emph{with} $\lim_{x \rightarrow +\infty} q(x) = +\infty$, \emph{which are both determinable only on the basis of the knowledge of} $\mu_0$, \emph{and such that}
\begin{equation}\label{eq:doob}
\intethree |\vb|^2 q(|\vb|) \mu(\ud\vb, t) \leq C(\mu_0)
\end{equation}
\emph{holds true for all} $t \geq 0$, \emph{leading to}
\begin{equation}\label{eq:unifintcutoff}
\lim_{R \rightarrow +\infty} \sup_{t \geq 0} \int_{|\vb| \geq R} |\vb|^2 \mu(\ud \vb, t) = 0\ .
\end{equation}
\end{thm}
To comment on the statements of Theorem \ref{thm:cutoff}, it is worth recalling that the Wild sum was introduced in \cite{wil,mck7}, while up-to-date references are \cite{cl,puto} and Chapter 2D of \cite{vil}. Preservation of momentum and kinetic energy, being a cornerstone in classical kinetic theory, goes back to Boltzmann himself. Equation (\ref{eq:ViVj}) can be thought of as a weak form of \emph{propagation of chaos}, as well as a macroscopic version of the principle of \emph{equipartition of the energy}. Lastly, the relevance of (\ref{eq:doob})-(\ref{eq:unifintcutoff}) -- remarked in \cite{cl} also from a physical standpoint -- is here confirmed and proved, for the first time, in a very general setting. The key element of the proof is a probabilistic representation of the solution $\mut$ which has been recently introduced in \cite{doreACTA}. Here is only a short presentation of both ideas and notation, the reader being referred to Subsection 1.5 of the aforementioned work. The core of the probabilistic representation, which is valid only upon assuming (\ref{eq:cutoff}), is encapsulated in the identity
\begin{equation} \label{eq:main}
\hat{\mu}(\rho \ub, t) = \textsf{E}_t \left[e^{i\rho S(\ub)}\right] \ \ \ \ \ (\rho \in \rone, \ub \in S^2, t \geq 0)
\end{equation}
where $S(\ub)$ is a random sum of weighted random variables and $\et$ is an expectation, for every $t \geq 0$. To define these two objects in a precise way, consider the sample space $\Omega := \mathbb{N} \times \mathbb{T} \times [0, \pi]^{\infty} \times (0, 2\pi)^{\infty} \times (\rthree)^{\infty}$ endowed with the $\sigma$-algebra $\mathscr{F} := 2^{\mathbb{N}} \otimes  2^{\mathbb{T}} \otimes  \mathscr{B}([0, \pi]^{\infty}) \otimes  \mathscr{B}((0, 2\pi)^{\infty}) \otimes  \mathscr{B}((\rthree)^{\infty})
$ where, for any topological set $X$, $X^{\infty}$ is the set of all sequences $(x_1, x_2, \dots)$ with elements in $X$, $2^X$ is the power set and $\mathscr{B}(X)$ the Borel class on $X$. Then, $\mathbb{T} := \textsf{X}_{\substack{n \geq 1}} \mathbb{T}(n)$ and $\mathbb{T}(n)$ is the (finite) set of all \emph{McKean binary trees} with $n$ leaves, whose generic element will be indicated as
$\treen$. Denoting by $\nu,\ \{\tau_n\}_{n \geq 1},\ \{\phi_n\}_{n \geq 1},\ \{\vartheta_n\}_{n \geq 1}, \{\mathbf{V}_n\}_{n \geq 1}$ the coordinate random variables of $\Omega$, one can consider, for any $t \geq 0$, the unique probability distribution (p.d.) $\pt$ on $(\Omega, \mathscr{F})$ which makes these random elements stochastically independent, consistently with the following marginal p.d.'s:
\begin{enumerate}
\item[a)] $\pt[ \nu = n ] = e^{-t}(1 - e^{-t})^{n-1}$ for $n = 1, 2, \dots$, with the proviso that $0^0 := 1$.
\item[b)] $\{\tau_n\}_{n \geq 1}$ is a Markov sequence driven by the initial condition $\pt[\tau_1 = \tree_1] = 1$ and the transition probabilities
$$
\begin{array}{lll}
\pt[ \tau_{n+1} = \treenk \ | \ \tau_n = \treen ] &= \frac{1}{n} & \ \text{for}\ k = 1, \dots, n \\
\pt[ \tau_{n+1} = \tree_{n+1} \ | \ \tau_n = \treen ] &= 0 & \ \text{if}\ \tree_{n+1} \not\in \mathbb{G}(\treen)
\end{array}
$$
where, for a given $\treen$, $\treenk$ indicates the germination of $\treen$ at its $k$-th leaf, obtained by appending a two-leaved tree to the $k$-th leaf of $\treen$, and $\mathbb{G}(\treen)$ is the subset of $\mathbb{T}(n+1)$ containing all the germinations of $\treen$.
\item[c)] The elements of $\{\phi_n\}_{n \geq 1}$ are independent and identically distributed (i.i.d.) random numbers with p.d. $\beta(\ud \varphi) := \frac{1}{2} b(\cos\varphi) \sin\varphi \ud \varphi$, with $\varphi \in [0, \pi]$.
\item[d)] The elements of $\{\vartheta_n\}_{n \geq 1}$ are i.i.d. with uniform p.d. on $(0, 2\pi)$, $u_{(0, 2\pi)}$.
\item[e)] The elements of $\{\mathbf{V}_n\}_{n \geq 1}$ are i.i.d. with p.d. $\mu_0$, the initial datum for (\ref{eq:boltzmann}).
\end{enumerate}
Therefore, $\et$ is defined as the expectation w.r.t. $\pt$. As for $S(\ub)$, consider the array
$\boldsymbol{\pi} := \{\pi_{j, n}\ | \ j = 1, \dots, n; n \in \mathbb{N}\}$ of $[-1,1]$-valued random numbers obtained by setting
$\pi_{j, n} := \pi_{j, n}^{\ast}(\tau_n, (\phi_1, \dots, \phi_{n-1}))$ for $j = 1, \dots, n$ and $n$ in $\mathbb{N}$. The $\pi_{j, n}^{\ast}$'s are functions on $\mathbb{T}(n) \times [0, \pi]^{n-1}$ given by $\pi_{1, 1}^{\ast} \equiv 1$ and, for $n \geq 2$,
$$
\pi_{j, n}^{\ast}(\treen, \boldsymbol{\varphi}) := \left\{ \begin{array}{ll}
\pi_{j, n_l}^{\ast}(\mathfrak{t}_{n}^{l}, \boldsymbol{\varphi}^l) \cos\varphi_{n-1} & \text{for} \ j = 1, \dots, n_l \\
\pi_{j - n_l, n_r}^{\ast}(\mathfrak{t}_{n}^{r}, \boldsymbol{\varphi}^r) \sin\varphi_{n-1} & \text{for} \ j = n_l + 1, \dots, n
\end{array} \right.
$$
for every $\boldsymbol{\varphi} = (\boldsymbol{\varphi}^l, \boldsymbol{\varphi}^r, \varphi_{n-1})$ in $[0, \pi]^{n-1}$, with
$\boldsymbol{\varphi}^l := (\varphi_1, \dots, \varphi_{n_l-1})$ and $\boldsymbol{\varphi}^r := (\varphi_{n_l}, \dots, \varphi_{n-2})$ where $\treen^l$ and $\treen^r$ symbolize the two trees, of $n_l$ and $n_r$ leaves respectively, obtained by deleting the root node  of $\treen$. Apropos of the $\pi_{j, n}$'s, it is worth noting that
\begin{equation} \label{eq:sumpijn}
\sum_{j=1}^{n} \pi_{j, n}^{2} = 1
\end{equation}
holds for every $n$ in $\mathbb{N}$. Another constituent of the desired representation is the array $\boldsymbol{\mathrm{O}} := \{\mathrm{O}_{j, n}\ | \ j = 1, \dots, n; n \in \mathbb{N}\}$ of random matrices $\mathrm{O}_{j, n}$, taking values in the Lie group $\mathbb{SO}(3)$ of orthogonal matrices with positive determinant, defined by $\mathrm{O}_{j, n} := \mathrm{O}_{j, n}^{\ast}(\tau_n, (\phi_1, \dots, \phi_{n-1}), (\vartheta_1, \dots, \vartheta_{n-1}))$ for $j = 1, \dots, n$ and $n$ in $\mathbb{N}$. The $\mathrm{O}_{j, n}^{\ast}$'s are $\mathbb{SO}(3)$-valued functions obtained by putting $\mathrm{O}_{1, 1}^{\ast} \equiv \mathrm{Id}_{3 \times 3}$ and, for $n \geq 2$,
\begin{eqnarray}
&{}& \mathrm{O}_{j, n}^{\ast}(\treen, \boldsymbol{\varphi}, \boldsymbol{\theta}) \nonumber \\
&:=& \left\{ \begin{array}{ll}
\mathrm{M}^l(\varphi_{n-1}, \theta_{n-1}) \mathrm{O}_{j, n_l}^{\ast}(\treen^l, \boldsymbol{\varphi}^l, \boldsymbol{\theta}^l) & \text{for} \ j = 1, \dots, n_l \\
\mathrm{M}^r(\varphi_{n-1}, \theta_{n-1}) \mathrm{O}_{j - n_l, n_r}^{\ast}(\treen^r, \boldsymbol{\varphi}^r, \boldsymbol{\theta}^r) & \text{for} \ j = n_l + 1, \dots, n
\end{array} \right. \nonumber
\end{eqnarray}
for every $\treen$ in $\mathbb{T}(n)$, $\boldsymbol{\varphi}$ in $[0, \pi]^{n-1}$ and $\boldsymbol{\theta}$ in $(0, 2\pi)^{n-1}$. Here, $\boldsymbol{\theta}^l := (\theta_1, \dots, \theta_{n_l-1})$ and $\boldsymbol{\theta}^r := (\theta_{n_l}, \dots, \theta_{n-2})$ and, finally,
$$
\mathrm{M}^l(\varphi, \theta) := \left( \begin{array}{ccc} -\cos\theta \cos\varphi & \sin\theta & \cos\theta \sin\varphi \\
- \sin\theta \cos\varphi & -\cos\theta & \sin\theta \sin\varphi  \\
\sin\varphi & 0 & \cos\varphi \\
\end{array} \right)
$$
$$
\mathrm{M}^r(\varphi, \theta) := \left( \begin{array}{ccc} \sin\theta & \cos\theta \sin\varphi & -\cos\theta \cos\varphi \\
-\cos\theta & \sin\theta \sin\varphi & - \sin\theta \cos\varphi \\
0 & \cos\varphi & \sin\varphi \\
\end{array} \right) \ .
$$
As a final step, choose a non-random measurable function $\mathrm{B}$ from $S^2$ onto $\mathbb{SO}(3)$ such that $\mathrm{B}(\ub) \mathbf{e}_3 = \ub$ for every $\ub$ in $S^2$, and define the random functions $\psib_{j, n} : S^2 \rightarrow S^2$ through the relation $\psib_{j, n}(\ub) := \mathrm{B}(\ub) \mathrm{O}_{j, n} \mathbf{e}_3$ for $j = 1, \dots, n$ and $n$ in $\mathbb{N}$, with $\mathbf{e}_3 := (0, 0, 1)^t$, to get
\begin{equation} \label{eq:Su}
S(\ub) := \sum_{j = 1}^{\nu} \pi_{j, \nu} \psib_{j, \nu}(\ub) \cdot \mathbf{V}_j\ .
\end{equation}

\subsection{Main results} \label{sect:main}

There are now the elements to state the new results, condensed in
\begin{thm}\label{thm:main}
\emph{Let} $b$ \emph{satisfy} (\ref{eq:vwcutoff}) \emph{and let} $\mu_0$ \emph{be a Borel p.m. on} $\rthree$ \emph{such that}
$\mathfrak{m}_2 := \intethree |\vb|^2 \mu_0(\ud \vb) < +\infty$. \emph{Then, there exists a solution} $\{\mut\}_{t \geq 0}$ \emph{which meets Definition} \ref{defin:solution} \emph{with initial datum} $\mu_0$, \emph{which can be obtained as follows: After defining} $B_n := \int_{0}^{1} [b(x) \wedge n] \ud x$ \emph{and} $\{\mu_n(\cdot, t)\}_{t \geq 0}$ \emph{as the unique solution of} (\ref{eq:boltzmann}) \emph{with} $[b(x) \wedge n]/B_n$ \emph{and} $\mu_0$ \emph{as collision kernel and initial datum, respectively, there is a subsequence} $\{n_l\}_{l \geq 1}$ \emph{of integers such that} $\mu_{n_l}(\cdot, B_{n_l} t) \Rightarrow \mut$ \emph{for all} $t \geq 0$. \emph{Moreover, if} $R$ \emph{and} $f_R$ \emph{are as in Lemma} \ref{lm:extendedQ}, \emph{then} $\{\mu(\cdot, t) \circ f_{R}^{-1}\}_{t \geq 0}$ \emph{is a solution of} (\ref{eq:boltzmann}) \emph{with} $\mu_0 \circ f_{R}^{-1}$ \emph{as initial datum. Finally,} (\ref{eq:Vu0})-(\ref{eq:unifintcutoff}) \emph{continue to be valid with the same} $C(\mu_0)$ \emph{and} $q$ \emph{as in Theorem} \ref{thm:cutoff}.
\end{thm}
As recalled in the introduction, the main points of this theorem can be found in various works, which prove them under somewhat different hypotheses. Actually, some papers consider them as folklore. It is worth stressing that the existence of a solution $\mut$ as limit of $\mu_n(\cdot, B_n t)$ was proposed and proved by Arkeryd \cite{ark} as far as the Boltzmann equation with hard potentials, by means of entropy methods. An adaptation of the Arkeryd strategy to the same context of Theorem \ref{thm:main} has recently appeared in \cite{morimoto}. The uniqueness for this kind of solutions has been proved in \cite{tovil}, as far as the weak cutoff case, while it remains an open problem in the very weak cutoff framework. Indeed, an attempt to fill this gap is contained in \cite{morimoto}, but that proof suffers the same drawback which has been mentioned in the introduction, so that further work is required to complete the argument. Finally, the statement of (\ref{eq:Vu0})-(\ref{eq:unifintcutoff}) in the general case of Maxwellian kernels satisfying (\ref{eq:vwcutoff}) represents, at the best of the author's knowledge, a novelty of this study which, in any case, gives a genuine physical meaning to these new solutions. Besides, if the hypotheses of Theorem 1 in \cite{doreACTA} are fulfilled, then the conclusion of that theorem, encapsulated in (16), remains valid, i.e. these new solutions converge to the Maxwellian distribution having the same momentum and energy of the initial datum, with the optimal time-rapidity w.r.t. the total variation distance.

\section{Proofs} \label{sect:proofs}

Gathered here are the proofs of Lemmata \ref{lm:villani} and \ref{lm:extendedQ} and of Theorems \ref{thm:cutoff} and \ref{thm:main}.

\subsection{Proof of Lemma \ref{lm:villani}} \label{sect:villani}

Observing that
\begin{eqnarray}
\vb_{\ast}(x) &=& \vb + |\wb - \vb| x \left[\sqrt{1 - x^2}(\cos\theta \ab(\ub) + \sin\theta \bb(\ub)) + x\ub\right]\nonumber \\
\wb_{\ast}(x) &=& \wb - |\wb - \vb| x \left[\sqrt{1 - x^2}(\cos\theta \ab(\ub) + \sin\theta \bb(\ub)) + x\ub\right] \nonumber
\end{eqnarray}
hold for every $\vb \neq \wb$ and $x \in (-1, 1)$, one gets
\begin{eqnarray}
\frac{\ud\vb_{\ast}}{\ud x} &=& - \frac{\ud\wb_{\ast}}{\ud x} = |\wb - \vb| \left[\frac{1 - 2x^2}{(1 - x^2)^{1/2}}(\cos\theta \ab(\ub) + \sin\theta \bb(\ub)) + 2x\ub\right]\nonumber \\
\frac{\ud^2\vb_{\ast}}{\ud x^2} &=& - \frac{\ud^2\wb_{\ast}}{\ud x^2} = |\wb - \vb| \left[\frac{-3x + 2x^3}{(1 - x^2)^{3/2}}(\cos\theta \ab(\ub) + \sin\theta \bb(\ub)) + 2\ub\right]\ . \nonumber
\end{eqnarray}
Whence,
\begin{eqnarray}
\Big{|}\frac{\ud\vb_{\ast}}{\ud x}\Big{|}^2 &=& \Big{|}\frac{\ud\wb_{\ast}}{\ud x}\Big{|}^2 = |\wb - \vb|^2 \left[
\frac{(1 - 2x^2)^2}{1 - x^2} + 4x^2\right]\nonumber \\
\Big{|}\frac{\ud^2\vb_{\ast}}{\ud x^2}\Big{|} &=& \Big{|}\frac{\ud^2\wb_{\ast}}{\ud x^2}\Big{|} = |\wb - \vb| \left[
\frac{(-3x + 2x^3)^2}{(1 - x^2)^3} + 4\right]^{1/2}\ . \nonumber
\end{eqnarray}
At this stage, for the first derivatives one has
\begin{eqnarray}
&& \intethree \intethree \int_{-1}^{1}\int_{0}^{2\pi} \int_{0}^{1} \left[\Big{|}\frac{\ud \vb_{\ast}}{\ud x}(s \xi)\Big{|}^2 + \Big{|}\frac{\ud \wb_{\ast}}{\ud x}(s \xi)\Big{|}^2\right] \ind\{\vb \neq \wb\} (1-s) \xi^2 b(\xi) \times \nonumber \\
&\times&  \ud s\ud\theta\ud\xi \chi(\ud \vb) \chi(\ud \wb)
\leq 2\pi\intethree \intethree |\wb - \vb|^2 \chi(\ud \vb) \chi(\ud \wb) \int_{-1}^{1}\xi^2 b(\xi)\ud\xi \times\nonumber\\
&\times& \int_{0}^{1}(1-s) \sup_{\xi \in [0,1]}\left(\frac{(1 - 2s^2\xi^2)^2}{1 - s^2\xi^2} + 4s^2\xi^2\right) \ud s\nonumber
\end{eqnarray}
the RHS being finite in view of (\ref{eq:vwcutoff}), the condition $\intethree |\vb|^2 \chi(\ud \vb) < +\infty$ and the fact that, for every $s$ in $(0,1)$,
$$
(1-s) \sup_{\xi \in [0,1]}\left(\frac{(1 - 2s^2\xi^2)^2}{1 - s^2\xi^2} + 4s^2\xi^2\right) \leq 14 \ .
$$
Analogously, for the second derivatives one has
\begin{eqnarray}
&& \intethree \intethree \int_{-1}^{1}\int_{0}^{2\pi} \int_{0}^{1} \left[\Big{|}\frac{\ud^2 \vb_{\ast}}{\ud x^2}(s \xi)\Big{|} + \Big{|}\frac{\ud^2 \wb_{\ast}}{\ud x^2}(s \xi)\Big{|}\right] \ind\{\vb \neq \wb\}(1-s) \xi^2 b(\xi) \times \nonumber \\
&\times& \ud s\ud\theta\ud\xi \chi(\ud \vb) \chi(\ud \wb) \leq 2\pi\intethree \intethree |\wb - \vb| \chi(\ud \vb) \chi(\ud \wb) \int_{-1}^{1}\xi^2 b(\xi)\ud\xi \times\nonumber\\
&\times& \int_{0}^{1}(1-s) \sup_{\xi \in [0,1]}\left(\frac{(-3s\xi + 2s^3\xi^3)^2}{(1 - s^2\xi^2)^3} + 4\right)^{1/2} \ud s\nonumber
\end{eqnarray}
and again the RHS is finite
in view of (\ref{eq:vwcutoff}), the condition $\intethree |\vb|^2 \chi(\ud \vb) < +\infty$ and the fact that, for every $s$ in $(0,1)$,
$$
(1-s) \sup_{\xi \in [0,1]}\left(\frac{(-3s\xi + 2s^3\xi^3)^2}{(1 - s^2\xi^2)^3} + 4\right)^{1/2} \leq \frac{2\sqrt{13}}{\sqrt{1-s}} + 2 \ .
$$

\subsection{Proof of Lemma \ref{lm:extendedQ}} \label{sect:proofextendedQ}

The following facts are valid for every $\omb \in S^2$: $(\vb, \wb) \mapsto (\vb_{\ast}, \wb_{\ast})$ is a linear diffeomorphism of $\rone^6$ with determinant identically equal to 1, and $b\left(\frac{\wb - \vb}{|\wb - \vb|} \cdot \omb \right) = b\left(\frac{\wb_{\ast} - \vb_{\ast}}{|\wb_{\ast} - \vb_{\ast}|} \cdot \omb \right)$ is valid for every $(\vb, \wb) \in \rone^6\setminus\{\vb = \wb\}$. Then, under (\ref{eq:cutoff}), $(\vb, \wb, \omb) \mapsto p(\vb_{\ast}) q(\wb_{\ast}) b\left(\frac{\wb - \vb}{|\wb - \vb|} \cdot \omb \right)$ is summable on $\rone^6 \times S^2$, with integral on the whole domain equal to 1 for any couple $(p,q)$ of probability densities on $\rthree$, by virtue of Fubini and Tonelli's theorems. Hence, $Q[p,q](\cdot)$ itself is a well-defined probability density function on $\rthree$. The change of variable $(\vb, \wb) \leftrightarrow (\vb_{\ast}, \wb_{\ast})$ gives, for all $\psi \in \Cb$,
$$
\intethree \psi(\vb) Q[p,q](\vb) \ud\vb = \intethree\intethree\int_{S^2}\psi(\vb_{\ast}) b\left(\frac{\wb - \vb}{|\wb - \vb|} \cdot \omb \right) p(\vb)q(\wb) \unifSo \ud\vb \ud\wb
$$
and now the core of the proof hinges on the fact that
$$
H_{\psi} : (\vb, \wb) \mapsto \left\{\begin{array}{ll}
\int_{S^2} \psi(\vb_{\ast}) b\left(\frac{\wb - \vb}{|\wb - \vb|} \cdot \omb \right) \unifSo
& \text{if} \ \vb \neq \wb \\
\psi(\vb) & \text{if} \ \vb = \wb
\end{array} \right.
$$
is bounded and continuous on $\rone^6$. Indeed, the continuity at some $(\vb_0, \vb_0)$ is obvious and can be checked directly.
Otherwise, take two linearly independent unitary vectors $\ub_{1}^{\bot}$ and $\ub_{1}^{\bot}$ and choose two orthonormal bases $\{\mathbf{a}_1(\ub), \mathbf{b}_1(\ub), \ub\}$ and $\{\mathbf{a}_2(\ub), \mathbf{b}_2(\ub), \ub\}$ of $\rthree$ whose components $\mathbf{a}_{i}(\ub)$ and $\mathbf{b}_i(\ub)$ vary with continuity w.r.t. $\ub \in S^2\setminus\{\pm \ub_{i}^{\bot}\}$, $i = 1,2$ respectively. Changing the coordinate in the spherical integral as in Subsection \ref{sect:weak} yields
\begin{eqnarray}
H_{\psi}(\vb, \wb) &=& \frac{1}{4\pi}\int_{0}^{2\pi}\int_{0}^{\pi} \psi\Big{(}\vb + |\wb-\vb|\cos\varphi\Big{[}\cos\theta\sin\varphi \mathbf{a}_{i}\left(\frac{\wb - \vb}{|\wb - \vb|}\right) \nonumber \\
&+& \sin\theta\sin\varphi \mathbf{b}_{i}\left(\frac{\wb - \vb}{|\wb - \vb|}\right) + \cos\varphi\frac{\wb - \vb}{|\wb - \vb|}\Big{]}\Big{)} b(\cos\varphi)\sin\varphi \ud\varphi\ud\theta \nonumber
\end{eqnarray}
which shows, by the dominated convergence theorem, that the set of the eventual discontinuities of $H_{\psi}$ is contained in both the subspaces $\{(\vb, \wb) \in \mathbb{R}^6 \ | \ \vb - \wb \in \ell_i\}$, $i = 1,2$, where $\ell_i := \{\lambda \ub_{i}^{\bot}\ |\ \lambda \in \rone\}$. Since $\ell_1 \cap \ell_2 = \{\mathbf{0}\}$, the eventual discontinuities are contained in $\{(\vb, \wb) \in \mathbb{R}^6 \ | \ \vb = \wb\}$, which was already known to be formed of points of continuity. Thus, the claim about $H_{\psi}$ is verified and, consequently, $\intethree \psi(\vb) Q[p_n,q_n](\vb) \ud\vb = \intethree\intethree H_{\psi}(\vb, \wb) \zeta_n(\ud\vb)\eta_n(\ud\wb)$ converges to $\intethree\intethree H_{\psi}(\vb, \wb) \zeta(\ud\vb)\eta(\ud\wb)$, as $n$ goes to infinity. To conclude that the limit is of the form $\intethree \psi(\vb) \mathcal{Q}[\zeta,\eta](\ud\vb)$ for some specific Borel p.m. $\mathcal{Q}[\zeta,\eta]$ on $\rthree$, it is enough to choose $\psi(\vb) = e^{i \xib \cdot \vb}$ with $\xib \neq \mathbf{0}$ and invoke the L\'{e}vy continuity theorem. Indeed, $\intethree\intethree H_{\psi}(\vb, \wb) \zeta(\ud\vb)\eta(\ud\wb)$ turns out to be a continuous function of $\xib$ in this case, by the dominated convergence theorem, since $H_{\psi}$ is continuous w.r.t. $\xib$ and $|H_{\psi}| \leq 1$. Consequently, the Fourier transform of the density $Q[p_n,q_n](\cdot)$ converges pointwise to a continuous function $g$, necessarily of the form $g(\xib) = \intethree e^{i \xib \cdot \vb} \mathcal{Q}[\zeta,\eta](\ud\vb)$. In this way, $\mathcal{Q}[\zeta, \eta]$ is well-defined and (\ref{eq:bobylevQ}) holds true. As for (\ref{eq:truccoR}), the weak continuity of $\mathcal{Q}$ w.r.t. $(\zeta, \eta)$ reduces the problem to checking the validity of $Q[p(S \cdot), q(S \cdot)](\vb) = Q[p(\cdot), q(\cdot)](S \vb)$ for every orthogonal matrix $S \in \mathbb{O}(3)$. The considerations at the beginning of this subsection entail
\begin{eqnarray}
Q[p(S \cdot), q(S \cdot)](\vb) &=& \intethree\int_{S^2} p(S \vb_{\ast})q(S \wb_{\ast}) b\left(\frac{\wb - \vb}{|\wb - \vb|} \cdot \omb \right) \unifSo \ud \wb\nonumber \\
&=& \intethree\int_{S^2} p(S\vb + [(S\wb - S\vb) \cdot \omb] \ \omb) \times \nonumber \\
&\times& q(S\wb - [(S\wb - S\vb) \cdot \omb] \ \omb) b\left(\frac{S\wb - S\vb}{|S\wb - S\vb|} \cdot \omb \right) \unifSo \ud \wb\nonumber \\
&=& \intethree\int_{S^2} p(S\vb + [(\zb - S\vb) \cdot \omb] \ \omb) \times \nonumber \\
&\times& q(\zb - [(\zb - S\vb) \cdot \omb] \ \omb) b\left(\frac{\zb - S\vb}{|\zb - S\vb|} \cdot \omb \right) \unifSo \ud \zb\nonumber \\
&=& Q[p(\cdot), q(\cdot)](S \vb)\nonumber
\end{eqnarray}
that is the desired identity.

Finally, to prove (\ref{eq:bobylev}), consider again $H_{\psi}$ with $\psi(\vb) = e^{i \xib \cdot \vb}$ and $\xib \neq \mathbf{0}$. An application the change of variable $\omb \leftrightarrow R \omb$, where $R \in \mathbb{O}(3)$ is such that $R^t \frac{\wb - \vb}{|\wb - \vb|} = \frac{\xib}{|\xib|}$, shows that $H_{\psi}(\vb, \wb) = e^{i \xib \cdot \vb} \int_{S^2} e^{i (\xib \cdot \omb)[(\wb - \vb) \cdot \omb]} b\left(\frac{\xib}{|\xib|} \cdot \omb\right)$.

\subsection{Proof of Theorem \ref{thm:cutoff} (existence and uniqueness)}

Since $\sum_{n = 1}^{+\infty} e^{-t} (1 - e^{-t})^{n-1} = 1$ for every $t \in [0, +\infty)$, the sum (\ref{eq:wild}) is a mixture, and so a well-defined Borel p.m. on $\rthree$ which coincides with $\mu_0$ when $t=0$. As for point ii) of Definition \ref{defin:solution}, consider $F_{\psi}(t) := \sum_{n = 1}^{+\infty} e^{-t} (1 - e^{-t})^{n-1} \intethree \psi(\vb) \mathcal{Q}_n[\mu_0](\ud\vb)$ with $\psi \in \Cb$ not identically zero. Putting $a_n := \intethree \psi(\vb) \mathcal{Q}_n[\mu_0](\ud\vb)$, the radius of convergence of the power series $\sum_{n = 0}^{+\infty} a_{n+1} z^n$ is at least 1, since $\limsup_{n\rightarrow \infty} \sqrt[n]{|a_{n+1}|} \leq \limsup_{n\rightarrow \infty} \sqrt[n]{\lnorm \psi \rnorm_{\infty}} = 1$, with $\lnorm \psi \rnorm_{\infty} := \sup_{\vb \in \rthree} |\psi(\vb)|$. Thus, the convergence of the series is uniform when $t \in [-\log(1+M), -\log(1-M)]$, for every $M \in (0,1)$, and $F_{\psi}$ is analytic in $(-\log 2, +\infty)$. To prove that (\ref{eq:wild}) satisfies (\ref{eq:wboltzmann}), differentiate $F_{\psi}$ term-to-term and resort to (\ref{eq:iterationQ}) to obtain
\begin{eqnarray}
\frac{\ud}{\ud t} F_{\psi}(t) &=& -F_{\psi}(t) + \sum_{n = 1}^{+\infty} n e^{-2t} (1 - e^{-t})^{n-1} \intethree \psi(\vb) \mathcal{Q}_{n+1}[\mu_0](\ud\vb) \nonumber \\
&=& -F_{\psi}(t) + \sum_{n = 1}^{+\infty} \sum_{k = 1}^{n} e^{-2t} (1 - e^{-t})^{n-1} \intethree \psi(\vb)
\mathcal{Q}\left[\mathcal{Q}_{k}[\mu_0], \mathcal{Q}_{n+1-k}[\mu_0]\right](\ud\vb) \ . \nonumber
\end{eqnarray}
Moreover, $-F_{\psi}(t)$ coincides, by definition, with
\begin{eqnarray}
&& \sum_{n = 1}^{+\infty} \sum_{m = 1}^{+\infty} e^{-2t} (1 - e^{-t})^{n+m-1} \frac{1}{2}\intethree \intethree \int_{S^2}
[- \psi(\vb) - \psi(\wb)] \ind\{\vb \neq \wb\} \times \nonumber \\
&\times& b\left(\frac{\wb - \vb}{|\wb - \vb|} \cdot \omb \right)
\unifSo \mathcal{Q}_{n}[\mu_0](\ud \vb)\mathcal{Q}_{m}[\mu_0](\ud\wb)\nonumber
\end{eqnarray}
while, by (\ref{eq:bobylevQ}), the other summand in the expression giving $\frac{\ud}{\ud t} F_{\psi}$ is equal to
\begin{eqnarray}
&& \sum_{n = 1}^{+\infty} \sum_{k = 1}^{n} e^{-2t} (1 - e^{-t})^{n-1} \intethree \intethree \int_{S^2}
\psi(\vb_{\ast}) \ind\{\vb \neq \wb\}\times \nonumber \\
&\times& b\left(\frac{\wb - \vb}{|\wb - \vb|} \cdot \omb \right)
\unifSo \mathcal{Q}_{k}[\mu_0](\ud \vb)\mathcal{Q}_{n+1-k}[\mu_0](\ud\wb)\nonumber\\
&=& \sum_{n = 1}^{+\infty} \sum_{m = 1}^{+\infty} e^{-2t} (1 - e^{-t})^{n+m-1}
\frac{1}{2}\intethree \intethree \int_{S^2}[\psi(\vb_{\ast}) + \psi(\wb_{\ast})] \ind\{\vb \neq \wb\} \times \nonumber \\
&\times& b\left(\frac{\wb - \vb}{|\wb - \vb|} \cdot \omb \right)
\unifSo \mathcal{Q}_{n}[\mu_0](\ud \vb)\mathcal{Q}_{m}[\mu_0](\ud\wb)\nonumber
\end{eqnarray}
which shows that point iv) of Definition \ref{defin:solution} is fulfilled.

The proof of uniqueness is based the fact that any weak solution $\overline{\mu}(\cdot, t)$ of (\ref{eq:boltzmann}) with initial datum $\mu_0$ must coincide with (\ref{eq:wild}), denoted by $\mut$. Indeed, under the validity of (\ref{eq:cutoff}), integrating (\ref{eq:wboltzmann}) w.r.t. $t$ shows that $\overline{\mu}(\cdot, t)$ must satisfy
\begin{eqnarray}
\intethree \psi(\vb) \overline{\mu}(\ud \vb, t) &=& e^{-t} \intethree \psi(\vb) \mu_0(\ud \vb) + \int_{0}^{t}
\intethree \intethree \int_{S^2} e^{-(t-s)} \psi(\vb_{\ast}) \ind\{\vb \neq \wb\} \times \nonumber \\
&\times& b\left(\frac{\wb - \vb}{|\wb - \vb|} \cdot \omb \right)
\unifSo \overline{\mu}(\ud \vb, s) \overline{\mu}(\ud \wb, s) \ud s \label{eq:wboltzmannintegral}
\end{eqnarray}
for every $t \in [0, +\infty)$ and $\psi \in \Cb$. The key point hinges on the inequality
\begin{equation}\label{eq:morgenstern}
\intethree \psi(\vb) \overline{\mu}(\ud \vb, t) \geq \sum_{n = 1}^{2^N} e^{-t} (1 - e^{-t})^{n-1} \intethree \psi(\vb) \mathcal{Q}_n[\mu_0](\ud\vb)
\end{equation}
which proves to be valid for every $t \in [0, +\infty)$, $N \in \mathbb{N}_0$ and every non-negative $\psi \in \Cb$. Arguing by induction, start by noticing that the case $N = 0$ is a direct consequence of (\ref{eq:iterationQ}) and (\ref{eq:wboltzmannintegral}). Then, recall from Subsection \ref{sect:proofextendedQ} that $(\vb,\wb) \mapsto \int_{S^2}\psi(\vb_{\ast}) b\left(\frac{\wb - \vb}{|\wb - \vb|} \cdot \omb \right) \unifSo$ is bounded and continuous and consider the latter term in the RHS of (\ref{eq:wboltzmannintegral}) with $\sum_{n = 1}^{2^N} e^{-s} (1 - e^{-s})^{n-1} \mathcal{Q}_n[\mu_0](\ud\vb)$ in place of $\overline{\mu}(\cdot, s)$, which turns out to coincide with
\begin{eqnarray}
&&\sum_{n = 1}^{2^N}\sum_{m = 1}^{2^N} \int_{0}^{t}e^{-(t-s)} e^{-2s} (1 - e^{-s})^{n+m-2} \ud s \intethree \intethree \int_{S^2}
\psi(\vb_{\ast}) \ind\{\vb \neq \wb\} \times \nonumber \\
&\times& b\left(\frac{\wb - \vb}{|\wb - \vb|} \cdot \omb \right)\unifSo \mathcal{Q}_{n}[\mu_0](\ud \vb)\mathcal{Q}_{m}[\mu_0](\ud\wb) \ . \nonumber
\end{eqnarray}
This expression is equal to $\sum_{n = 2}^{2^{N+1}} e^{-t} (1 - e^{-t})^{n-1}\intethree \psi(\vb) \mathcal{Q}_n[\mu_0](\ud\vb)$, thanks to (\ref{eq:bobylevQ}), the identity $\int_{0}^{t} e^{-(t-s)} e^{-2s} (1 - e^{-s})^{n+m-2} \ud s = \frac{1}{n+m-1}e^{-t} (1 - e^{-t})^{n+m-1}$ and (\ref{eq:iterationQ}). Thus, if (\ref{eq:morgenstern}) is valid for a certain $N \in \mathbb{N}_0$, then the RHS of (\ref{eq:wboltzmannintegral}) is not less than $\sum_{n = 1}^{2^{N+1}} e^{-t} (1 - e^{-t})^{n-1}\intethree \psi(\vb) \mathcal{Q}_n[\mu_0](\ud\vb)$, which proves the inequality at issue for every $N \in \mathbb{N}_0$. To conclude, observe that (\ref{eq:morgenstern}) entails $\overline{\mu}(B, t) \geq \mu(B, t)$ for every $t \in [0, +\infty)$ and $B \in \borelthree$, which is tantamount to asserting that $\overline{\mu}(\cdot, t)$ coincides with  $\mut$ for every $t \in [0, +\infty)$.

Finally, take $R$ and $f_R$ as in Lemma \ref{lm:extendedQ} and consider the solution of (\ref{eq:boltzmann}) with $\mu_0 \circ f_{R}^{-1}$ as initial datum. From (\ref{eq:wild}), this solution is given by $\sum_{n = 1}^{+\infty} e^{-t} (1 - e^{-t})^{n-1} \mathcal{Q}_n[\mu_0 \circ f_{R}^{-1}](\cdot)$, while $\mathcal{Q}_n[\mu_0 \circ f_{R}^{-1}] = \mathcal{Q}_n[\mu_0] \circ f_{R}^{-1}$ follows from (\ref{eq:truccoR}) and (\ref{eq:iterationQ}) by an obvious induction argument. Hence, the above sum coincides with $\{\mu(\cdot, t) \circ f_{R}^{-1}\}_{t \geq 0}$, where $\mut$ is the solution of (\ref{eq:boltzmann}) with initial datum $\mu_0$.

\subsection{Some preparatory results}

This subsection contains two technical lemmata, which will come in useful later on. The first statement consists in a refinement of a classical result about uniform integrability, whose original form is contained, e.g., in Section 7.VI of \cite{doob} or in Section 2.II of \cite{meyer}. The inspiration for the following refined version has come from the contents
of Section 3 of \cite{gtw} and from Lemma 2 of \cite{tovil}.
\begin{lm}\label{lm:doob}
\emph{Let} $\gamma$ \emph{be a Borel p.m. on} $[0,+\infty)$ \emph{such that} $\int_{0}^{+\infty} x\gamma(\ud x) < +\infty$. \emph{Then, there exists a function} $G : [0,+\infty) \rightarrow [0,+\infty)$, \emph{depending on} $\gamma$, \emph{with the following properties:}
\begin{enumerate}
\item[i)] $\int_{0}^{+\infty} G(x)\gamma(\ud x) < +\infty$;
\item[ii)] $G$ \emph{is strictly increasing and continuous, with} $G(0) = 0$;
\item[iii)] $\lim_{x \rightarrow +\infty} G(x)/x = +\infty$;
\item[iv)] \emph{there exists a discrete set} $\Delta \subset (0,+\infty)$ \emph{such that} $G \in \mathrm{C}^1((0,+\infty)\setminus \Delta)$;
\item[v)] \emph{there exists a constant} $\lambda_1 > 1$ \emph{such that} $G(x) \leq xG^{'}(x) \leq \lambda_1 G(x)$ \emph{for all} $x \in (0,+\infty)\setminus \Delta$;
\item[vi)] \emph{there exists a constant} $\lambda_2 > 1$ \emph{such that} $G^{'}(2x) \leq \lambda_2 G^{'}(x)$ \emph{for all}
$x \in (0,+\infty)\setminus \Delta$.
\end{enumerate}
\end{lm}

\noindent \emph{Proof}: Put $g(x) := \ind_{[0,1)}(x) + \sum_{n=0}^{\infty} A_n \ind_{[2^n,2^{n+1})}(x)$ for every $x \in [0, +\infty)$, where $\{A_n\}_{n \geq 0}$ is a suitable sequence of real numbers, to be determined from the knowledge of $\gamma$. General properties of this sequence must be the following:
\begin{enumerate}
\item[a)] $1 \leq A_n \leq A_{n+1}$ for all $n \in \mathbb{N}_0$;
\item[b)] $\lim_{n \rightarrow +\infty} A_n = +\infty$;
\item[c)] $\sup_{n \in \mathbb{N}_0} A_{n+1}/A_n < +\infty$.
\end{enumerate}
Define also $G(x) := \int_{0}^{x} g(y)\ud y$ for every $x \in [0, +\infty)$ and $\Delta := \{2^n\ |\ n \in \mathbb{N}_0\}$. At this stage, note that the above setting is enough to guarantee, independently of the specific determination of $\{A_n\}_{n \geq 0}$, the validity of the points from ii) to iv), as well as the inequality $G(x) \leq xG^{'}(x)$ for all $x \in (0,+\infty)\setminus \Delta$. Point vi) holds true after putting $\lambda_2 := \max\{A_0, \sup_{n \in \mathbb{N}_0} A_{n+1}/A_n\}$. As to the remaining inequality at point v), one shows that it is in force for all $x \in (0,+\infty)\setminus \Delta$ with $\lambda_1 := 2\lambda_2$. Indeed, when $x \in (0, 1)$ suffice it to know that $\lambda_1 > 1$. When $x \in (1, 2)$ the fact that $\lambda_1 \geq A_0$ yields $\lambda_1 G(x) - xg(x) \geq A_0(A_0 - 1)(x - 1) \geq 0$. When $x \in [2^m,2^{m+1})$ for some integer $m \in \mathbb{N}$, observe that the thesis is equivalent to the validity of $\lambda_1 \left[2^m A_m - 1 - \sum_{n=0}^{m-1} 2^n A_n\right] \leq A_m (\lambda_1 - 1) x$. Since the RHS is minimum when $x = 2^m$, it is enough to test this inequality in correspondence of this minimum point, reducing the problem to checking that $\sup_{m \geq 1} (2^m A_m)/(1 + \sum_{n=0}^{m-1} 2^n A_n) \leq \lambda_1$, which follows in view of
$$
\frac{2^m A_m}{1 + \sum_{n=0}^{m-1} 2^n A_n} \leq \frac{2^m \lambda_2 A_{m-1}}{2^{m-1} A_{m-1}} = 2\lambda_2\ .
$$
After showing that the validity of a)-c) entails that ii)-vi) are in force, the conclusion of the proof focuses on the specific determination of $\{A_n\}_{n \geq 0}$ in conformity with a)-c) and i). Accordingly, consider the distribution function $\Gamma(x) := \gamma([0, x])$, for every $x \in [0, +\infty)$, and define $\Gamma^{\ast}(x) := 1 - \Gamma(x)$. Next, integrate by parts to obtain $\int_{0}^{z} G(x)\gamma(\ud x) = -\Gamma^{\ast}(z) G(z) +  \int_{0}^{z} \Gamma^{\ast}(x) g(x) \ud x$ for all $z \in [0, +\infty)$. As for the latter integral, write
\begin{eqnarray}
\int_{0}^{z} \Gamma^{\ast}(x) g(x) \ud x &\leq& \int_{0}^{1} \Gamma^{\ast}(x)\ud x + \sum_{n=0}^{\infty} A_n \int_{2^n}^{2^{n+1}}
\Gamma^{\ast}(x)\ud x \nonumber \\
&\leq& 1 + \sum_{n=0}^{\infty} A_n \Big{(}\sum_{k = 2^n}^{2^{n+1}-1} \alpha_k \Big{)} = 1 + \sum_{k=1}^{\infty} A_{n(k)} \alpha_k \label{eq:meyer1}
\end{eqnarray}
where $\alpha_k := \Gamma^{\ast}(k)$ and $n(k)$ is the only integer such that $2^{n(k)} \leq k < 2^{n(k)+1}$. Then, choose a sequence of positive integers $\{r_n\}_{n \geq 1}$ such that $r_n \leq r_{n+1}$ and $\int_{r_n}^{+\infty} x \gamma(\ud x) \leq 2^{-n}$ for every $n \in \mathbb{N}$, which is possible by virtue of the hypothesis $\int_{0}^{+\infty} x\gamma(\ud x) < +\infty$. Whence, $2^{-n} \geq \int_{r_n}^{+\infty} x \gamma(\ud x) \geq \sum_{k=r_n}^{+\infty} k (\alpha_k - \alpha_{k+1}) \geq \sum_{k=r_n}^{+\infty} \alpha_k$, proving that the series $\sum_{n=1}^{\infty}\sum_{k=r_n}^{+\infty} \alpha_k$ is convergent.
By inverting the summation order, put the last series in the form $\sum_{n=1}^{+\infty} B_n \alpha_n$, with $B_n := \sum_{j=1}^{+\infty} \ind\{r_j \leq n\}$, which shows, in addition, that $B_n \leq B_{n+1}$ is in force for every $n \in \mathbb{N}$, and $\lim_{n \rightarrow +\infty} B_n = +\infty$. Next, introduce a new sequence $\{B_n^{\ast}\}_{n \geq 1}$ by setting $B_1^{\ast} := B_1$ and $B_n^{\ast} := \min\{B_m\ | \ B_m > B_n\}$, which satisfies $\lim_{n \rightarrow +\infty} B_n^{\ast} = +\infty$ and $n-1 \leq B_n^{\ast} < B_{n+1}^{\ast}$ for all $n \in \mathbb{N}$. With a view to the determination of $\{A_n\}_{n \geq 0}$, consider the function $h : [0,+\infty) \rightarrow [0,+\infty)$ defined by $h(0) = 0$, $h(n) = B_n^{\ast} + 1$ for all $n \in \mathbb{N}$ and by a linear interpolation in correspondence of the remaining values of $x$. This function turns out to be continuous and strictly increasing, and meets $h(x) \geq x$ for all $x \in [0,+\infty)$. Its inverse $h^{-1}$ is again strictly increasing, diverges at infinity and meets $h^{-1}(x) \leq x$ for all $x  \in [0,+\infty)$, so that one can finally put $A_0 = A_1$ and $A_n := h^{-1}(B_{n} + 1) + 1$, for every $n \in \mathbb{N}$. At this stage, points a)-b) are automatically satisfied while, as to point c), it suffices to observe that
$$
\sup_{n \in \mathbb{N}_0} \frac{A_{n+1}}{A_n} = \sup_{n \in \mathbb{N}_0} \frac{h^{-1}(B_{n+1}^{\ast} + 1) + 1}{h^{-1}(B_{n}^{\ast} + 1) + 1} = \sup_{n \in \mathbb{N}_0} \frac{n + 2}{n + 1} = 2\ .
$$
The validity of point i) follows from
$$
\sum_{k=1}^{\infty} A_{n(k)} \alpha_k \leq \sum_{k=1}^{\infty} A_k \alpha_k \leq \sum_{k=1}^{\infty} B_k \alpha_k + 2 \sum_{k=1}^{\infty} \alpha_k < +\infty
$$
which shows, via (\ref{eq:meyer1}), that $\int_{0}^{+\infty} \Gamma^{\ast}(x) g(x) \ud x < +\infty$. The conclusion ensues from the above-mentioned integration by parts, which gives $\int_{0}^{+\infty} G(x) \gamma(\ud x) \leq \int_{0}^{+\infty} \Gamma^{\ast}(x) g(x) \ud x$. \  $\square$

As a straightforward corollary of points ii)-vi) of this lemma, one can show that $G$ meets the following additional properties:
\begin{enumerate}
\item[i')] $G(2x) \leq \lambda_3 G(x)$ for all $x \in [0, +\infty)$, with $\lambda_3 := 2\lambda_2$;
\item[ii')] $G(x) = x \mathfrak{G}(x)$, where $\mathfrak{G} : [0, +\infty) \rightarrow [0, +\infty)$ is non-decreasing;
\item[iii')] $G(x) \leq G(1) x^{\lambda_1}$ for all $x \in [1, +\infty)$;
\item[iv')] $G(\sum_{i=1}^{2^m} x_i) \leq \lambda_3^m \sum_{i=1}^{2^m} G(x_i)$ for all $m \in \mathbb{N}$ and $x_1, \dots, x_{2^m} \in [0, +\infty)$.
\end{enumerate}
Indeed, i') follows from
$$
G(2x) = \int_{0}^{2x} G^{'}(y)\ud y = 2\int_{0}^{x} G^{'}(2y)\ud y \leq 2\lambda_2\int_{0}^{x} G^{'}(y)\ud y = \lambda_3G(x)\ .
$$
The next point ii') is an obvious consequence of $G(x) \leq xG^{'}(x)$. Then, iii') emanates by virtue of $xG^{'}(x) \leq \lambda_1 G(x)$ and iv') can be deduced, by means of an easy induction argument, from
$$
G(x_1 + x_2) \leq G(2\max\{x_1, x_2\}) \leq \lambda_3G(\max\{x_1, x_2\}) \leq \lambda_3[G(x_1) + G(x_2)]\ .
$$
Now, a close link between Lemma \ref{lm:doob} and the sum $S(\ub)$ appearing in (\ref{eq:Su}) is established by means of the following
\begin{lm}\label{lm:petrov}
\emph{Let the initial datum} $\mu_0$ \emph{satisfy} $\mathfrak{m}_2 := \intethree |\vb|^2 \mu_0(\ud \vb) < +\infty$. \emph{Then, there exists a positive constant} $C_1(\mu_0)$, \emph{depending only on} $\mu_0$, \emph{such that}
\begin{equation} \label{eq:petrovbound}
\et[G_{\ast}(S(\ub))] \leq C_1(\mu_0)
\end{equation}
\emph{is valid for every} $t \geq 0$ \emph{and} $\ub \in S^2$, \emph{where} $G_{\ast}(x) := G(x^2)$ \emph{and} $G$ \emph{is the same function provided by Lemma} \ref{lm:doob} \emph{when} $\gamma(A) := \intethree \ind\{|\vb|^2 \in A\} \mu_0(\ud\vb)$.
\end{lm}

\noindent \emph{Proof}: Since $\int_{0}^{+\infty} x\gamma(\ud x) = \mathfrak{m}_2$, the hypothesis in Lemma \ref{lm:doob} is fulfilled and $G$ is given accordingly. Then, introduce the $\sigma$-algebra $\mathscr{H} := \sigma\big{(}\nu,\ \{\tau_n\}_{n \geq 1},\ \{\phi_n\}_{n \geq 1},\\ \{\vartheta_n\}_{n \geq 1} \big{)}$ and invoke the structure of the probabilistic representation set forth in Subsection \ref{sect:grad} to have, for all $m \in \mathbb{N}$, $\ub \in S^2$ and $j = 1, \dots, \nu$,
\begin{eqnarray}
&& \et[G_{\ast}(2^m \pi_{j, \nu} \psib_{j, \nu}(\ub) \cdot \mathbf{V}_j)\ |\ \mathscr{H}] = \et[G(2^{2m} \pi_{j, \nu}^2 (\psib_{j, \nu}(\ub) \cdot \mathbf{V}_j)^2)\ |\ \mathscr{H}] \nonumber \\
&\leq& \lambda_3^{2m} \et[G(\pi_{j, \nu}^2 (\psib_{j, \nu}(\ub) \cdot \mathbf{V}_j)^2)\ |\ \mathscr{H}] \nonumber \\
&=& \lambda_3^{2m} \pi_{j, \nu}^2 \et[(\psib_{j, \nu}(\ub) \cdot \mathbf{V}_j)^2 \mathfrak{G}(\pi_{j, \nu}^2 (\psib_{j, \nu}(\ub) \cdot \mathbf{V}_j)^2)\ |\ \mathscr{H}] \nonumber \\
&\leq& \lambda_3^{2m} \pi_{j, \nu}^2 \et[|\mathbf{V}_j|^2 \mathfrak{G}(|\mathbf{V}_j|^2)] = \lambda_3^{2m} \pi_{j, \nu}^2 \et[G_{\ast}(|\mathbf{V}_j|)] \nonumber
\end{eqnarray}
since $|\psib_{j, \nu}(\ub) \cdot \mathbf{V}_j| \leq |\mathbf{V}_j|$. Thus, (\ref{eq:sumpijn}) entails
\begin{equation}\label{eq:petrovsomme}
\sum_{j = 1}^{\nu} \et[G_{\ast}(2^m \pi_{j, \nu} \psib_{j, \nu}(\ub) \cdot \mathbf{V}_j)\ |\ \mathscr{H}] \leq \lambda_3^{2m}
\int_{0}^{+\infty} G(x)\gamma(\ud x)
\end{equation}
for all $m \in \mathbb{N}$ and $\ub \in S^2$. After defining $G_{\ast, l}(x) := \min\{G_{\ast}(x), l\}$ for $l \in \mathbb{N}$
and checking that $\et[G_{\ast, l}(S(\ub))\ |\ \mathscr{H}] \leq l$, apply Lemma 2.4 in \cite{pe1} to obtain
\begin{equation}\label{eq:petrov2.4}
\et[G_{\ast, l}(S(\ub))\ |\ \mathscr{H}] = \int_{0}^{+\infty} \pt[|S(\ub)| \geq x\ |\ \mathscr{H}] \ud G_{\ast, l}(x)\ .
\end{equation}
Now, the conclusion hinges on the remark that $A \mapsto \pt[S(\ub) \in A\ |\ \mathscr{H}]$ is a (random) p.m. having the structure of probability law of a sum of independent random variables, which establishes a link with the subject of Chapter 2 of \cite{pe1} and allows the use of formula (2.33) therein, with $y = x 2^{-m}$, to get
\begin{eqnarray}
\pt[|S(\ub)| \geq x\ |\ \mathscr{H}] &\leq& \sum_{j = 1}^{\nu} \pt[|\pi_{j, \nu} \psib_{j, \nu}(\ub) \cdot \mathbf{V}_j| \geq x 2^{-m}\ |\ \mathscr{H}] \nonumber \\
&+& 2e^{2^m} \left(1 + \frac{x^2}{2^m \sum_{j = 1}^{\nu} \pi_{j, \nu}^2 \et[(\psib_{j, \nu}(\ub) \cdot \mathbf{V}_j)^2\ |\ \mathscr{H}]}\right)^{-2^m} \ . \nonumber
\end{eqnarray}
The combination of this last inequality with (\ref{eq:petrov2.4}) leads to the analysis of two terms, the former of which can be bounded as
\begin{eqnarray}
&& \sum_{j = 1}^{\nu} \int_{0}^{+\infty} \pt[|\pi_{j, \nu} \psib_{j, \nu}(\ub) \cdot \mathbf{V}_j| \geq x 2^{-m}\ |\ \mathscr{H}]\ud G_{\ast, l}(x) \nonumber \\
&\leq& \sum_{j = 1}^{\nu} \et[G_{\ast}(2^m \pi_{j, \nu} \psib_{j, \nu}(\ub) \cdot \mathbf{V}_j)\ |\ \mathscr{H}] \nonumber
\end{eqnarray}
for every $m, l \in \mathbb{N}$, which gives a finite upper bound thanks to (\ref{eq:petrovsomme}). As to the latter term, observe in advance that $\sum_{j = 1}^{\nu} \pi_{j, \nu}^2 \et[(\psib_{j, \nu}(\ub) \cdot \mathbf{V}_j)^2\ |\ \mathscr{H}] \leq \mathfrak{m}_2$ holds by means of (\ref{eq:sumpijn}) and the inequality $|\psib_{j, \nu}(\ub) \cdot \mathbf{V}_j| \leq |\mathbf{V}_j|$, so that property iii') of $G$ entails
\begin{eqnarray}
&& \int_{0}^{+\infty} \left(1 + \frac{x^2}{2^m \sum_{j = 1}^{\nu} \pi_{j, \nu}^2 \et[(\psib_{j, \nu}(\ub) \cdot \mathbf{V}_j)^2\ |\ \mathscr{H}]}\right)^{-2^m} \ud G_{\ast, l}(x) \nonumber \\
&\leq& \int_{0}^{+\infty} \left(1 + \frac{x^2}{2^m \mathfrak{m}_2}\right)^{-2^m} \ud G_{\ast}(x)
\leq G(1) \Big{[} 1 + 2\lambda_1 (2^m \mathfrak{m}_2)^{2^m} \int_{1}^{+\infty} x^{-2^{m+1} + 2\lambda_1 + 1}\ud x \Big{]} \nonumber
\end{eqnarray}
which gives a finite upper bound for every $l \in \mathbb{N}$, provided that $m$ is chosen in such a way that $-2^{m+1} + 2\lambda_1 + 1 < -1$. After putting $C_1(\mu_0) := \lambda_3^{2m} \int_{0}^{+\infty} G(x)\gamma(\ud x) + 2e^{2^m} G(1) \Big{[} 1 + \frac{\lambda_1 (2^m \mathfrak{m}_2)^{2^m}}{2^m - \lambda_1 - 1} \Big{]}$ with a suitable choice of $m$ (e.g., $m = [\log_2(\lambda_1 + 1)] + 1$, where $[x]$ denotes the integral part of $x$), one finally has $\et[G_{\ast, l}(S(\ub))\ |\ \mathscr{H}] \leq C_1(\mu_0)$ for every $l \in \mathbb{N}$, which entails (\ref{eq:petrovbound}).

\subsection{Proof of Theorem \ref{thm:cutoff} (evolution of the moments)} \label{sect:moments}

Throughout this subsection, assume that $\mathfrak{m}_2 := \intethree |\vb|^2 \mu_0(\ud \vb) < +\infty$ is in force and consider the sum $S(\ub)$ in (\ref{eq:Su}). A combination of Lyapunov and Cauchy-Schwartz's inequalities with (\ref{eq:sumpijn}) gives
\begin{eqnarray}
\et[(S(\ub))^2] &\leq& \et\Big{[}\nu \sum_{j = 1}^{\nu} \pi_{j, \nu}^2 (\psib_{j, \nu}(\ub) \cdot \mathbf{V}_j)^2\Big{]}
\leq \et\Big{[}\nu \sum_{j = 1}^{\nu} \pi_{j, \nu}^2 |\mathbf{V}_j|^2\Big{]}\nonumber \\
&=& \et\Big{[}\et\Big{[}\nu \sum_{j = 1}^{\nu} \pi_{j, \nu}^2 |\mathbf{V}_j|^2\ \big{|}\ \mathscr{G}\Big{]}\Big{]} = \et[\nu] \mathfrak{m}_2 \nonumber
\end{eqnarray}
for all $\ub \in S^2$, where $\mathscr{G} := \sigma\big{(}\nu,\ \{\tau_n\}_{n \geq 1},\ \{\phi_n\}_{n \geq 1} \big{)}$. Thus, taking account that $\et[\nu] = e^t$, the above set of inequalities shows that the first two moments of $S(\ub)$ are finite. To prove the former equality in (\ref{eq:Vu0}), observe that (\ref{eq:main}) entails $\et[S(\ub)] = \ub \cdot \intethree \vb \mu(\ud\vb, t)$ for every $t \geq 0$ and $\ub \in S^2$, and write
\begin{eqnarray}
\et[S(\ub)] &=& \et\Big{[}\et\Big{[}\sum_{j = 1}^{\nu} \pi_{j, \nu} \psib_{j, \nu}(\ub) \cdot \mathbf{V}_j\ \big{|}\ \mathscr{H}\Big{]}\Big{]} = \et\Big{[}\sum_{j = 1}^{\nu} \pi_{j, \nu} \psib_{j, \nu}(\ub)\Big{]} \cdot \overline{\mathbf{V}}
\nonumber \\
&=& \et\Big{[}\et\Big{[}\sum_{j = 1}^{\nu} \pi_{j, \nu} \psib_{j, \nu}(\ub)\ \big{|}\ \mathscr{G}\Big{]}\Big{]} \cdot \overline{\mathbf{V}} = \et\Big{[}\sum_{j = 1}^{\nu} \pi_{j, \nu}\et\Big{[} \psib_{j, \nu}(\ub)\ \big{|}\ \mathscr{G}\Big{]}\Big{]} \cdot  \overline{\mathbf{V}} \nonumber
\end{eqnarray}
with $\overline{\mathbf{V}} := \intethree \vb \mu_0(\ud \vb)$ and the same $\sigma$-algebra $\mathscr{H}$ used in the proof of Lemma \ref{lm:petrov}. Since $\et\Big{[}\psib_{j, \nu}(\ub)\ \big{|}\ \mathscr{G}\Big{]} = \pi_{j, \nu}\ub$ for every $\ub \in S^2$, in view of (111) in \cite{doreACTA}, invoke (\ref{eq:sumpijn}) to conclude that $\et\Big{[}\sum_{j = 1}^{\nu} \pi_{j, \nu}\et\Big{[} \psib_{j, \nu}(\ub)\ \big{|}\ \mathscr{G}\Big{]}\Big{]} = \ub$, which leads to the desired result. To prove the remaining relations, note that $\sum_{j = 1}^{\nu} \pi_{j, \nu} \psib_{j, \nu}(\ub) \cdot \overline{\mathbf{V}} = \ub \cdot \overline{\mathbf{V}}$, in view of the fact that $\delta_{\overline{\mathbf{V}}}$ is a stationary solution of (\ref{eq:boltzmann}),  to obtain, for all $\ub \in S^2$,
\begin{eqnarray}
\et[(S(\ub))^2] &=& \et\Big{[}\Big{(}\sum_{j = 1}^{\nu} \pi_{j, \nu} \psib_{j, \nu}(\ub) \cdot (\mathbf{V}_j - \overline{\mathbf{V}}) + \ub \cdot \overline{\mathbf{V}}\Big{)}^2\Big{]}\nonumber \\
&=& \et\Big{[}\Big{(}\sum_{j = 1}^{\nu} \pi_{j, \nu} \psib_{j, \nu}(\ub) \cdot (\mathbf{V}_j - \overline{\mathbf{V}})\Big{)}^2\Big{]} + (\ub \cdot \overline{\mathbf{V}})^2 \nonumber \\
&=& \et\Big{[}\sum_{j = 1}^{\nu} \pi_{j, \nu}^2 [\psib_{j, \nu}(\ub) \cdot (\mathbf{V}_j - \overline{\mathbf{V}})]^2\Big{]} + (\ub \cdot \overline{\mathbf{V}})^2 \ . \label{eq:secondmomSu}
\end{eqnarray}
An application of (187) in \cite{doreACTA} with $k = 2$ shows that, for all $\ub \in S^2$,
\begin{eqnarray}
&& \et\Big{[}\sum_{j = 1}^{\nu} \pi_{j, \nu}^2 [\psib_{j, \nu}(\ub) \cdot (\mathbf{V}_j - \overline{\mathbf{V}})]^2\Big{]} =
\frac{1}{3} \intethree |\vb - \overline{\mathbf{V}}|^2 \mu_0(\ud\vb) \nonumber \\
&+& \et\Big{[}\sum_{j = 1}^{\nu} \pi_{j, \nu}^2 \zeta_{j, \nu}\Big{]} \cdot \Big{[} \intethree \{(\ub \cdot (\vb - \overline{\mathbf{V}}))^2 - \frac{1}{3}|\vb - \overline{\mathbf{V}}|^2\} \mu_0(\ud\vb)\Big{]} \label{eq:secondmomSu1}
\end{eqnarray}
holds, where the $\zeta_{j, n}$'s are given by $\zeta_{j, n} := \zeta_{j, n}^{\ast}(\tau_n, (\phi_1, \dots, \phi_{n-1}))$ and the $\zeta_{j, n}^{\ast}$'s are defined on $\mathbb{T}(n) \times [0, \pi]^{n-1}$ by putting $\zeta_{1, 1}^{\ast} \equiv 1$ and, for $n \geq 2$,
$$
\zeta_{j, n}^{\ast}(\treen, \boldsymbol{\varphi}) := \left\{ \begin{array}{ll}
\zeta_{j, n_l}^{\ast}(\mathfrak{t}_{n}^{l}, \boldsymbol{\varphi}^l) \cdot (\frac{3}{2}\cos^2\varphi_{n-1} - \frac{1}{2})& \text{for} \ j = 1, \dots, n_l \\
\zeta_{j - n_l, n_r}^{\ast}(\mathfrak{t}_{n}^{r}, \boldsymbol{\varphi}^r) \cdot (\frac{3}{2}\sin^2\varphi_{n-1} - \frac{1}{2}) & \text{for} \ j = n_l + 1, \dots, n
\end{array} \right.
$$
for every $\boldsymbol{\varphi}$ in $[0, \pi]^{n-1}$. An application of the same techniques, contained in Appendix A.1 of \cite{doreACTA}, used to get (106) therein, shows that $\et\Big{[}\sum_{j = 1}^{\nu} \pi_{j, \nu}^2 \zeta_{j, \nu}\Big{]} = e^{-(1 - f_1(b)) t}$ for all $t \geq 0$, with $f_1(b) := \frac{3}{2}\int_{0}^{\pi} (\sin^4\varphi + \cos^4\varphi) \beta(\ud\varphi) - \frac{1}{2}$. It is interesting to observe that $-(1 - f_1(b)) = \frac{3}{2} \Lambda_b$, where $ \Lambda_b$ can be seen as the optimal rate of exponential convergence to equilibrium for the solutions of the SHBEMM. See Subsection 1.2 of \cite{doreACTA}. At this stage, the proof of the latter identity in (\ref{eq:Vu0}) follows easily from (\ref{eq:secondmomSu})-(\ref{eq:secondmomSu1}), which give
$$
\intethree |\vb|^2 \mu(\ud\vb, t) = \sum_{i=1}^{3} \et[(S(\mathbf{e}_i))^2] = \intethree |\vb - \overline{\mathbf{V}}|^2\mu_0(\ud\vb) + \sum_{i=1}^{3} (\mathbf{e}_i \cdot \overline{\mathbf{V}})^2 = \mathfrak{m}_2
$$
for all $t \geq 0$, where $\{\mathbf{e}_1, \mathbf{e}_2, \mathbf{e}_3\}$ is the canonical basis of $\rthree$. The proof of (\ref{eq:ViVj}), in the case $i = j$, is even more simpler, since it follows directly from the combination of (\ref{eq:secondmomSu})-(\ref{eq:secondmomSu1}) with $\ub = \mathbf{e}_i$. When $i \neq j$, start from the elementary remark that $\intethree v_iv_j \mu(\ud\vb, t)$ can be written as $\intethree \big{(}\frac{\sqrt{2}}{2}v_i + \frac{\sqrt{2}}{2}v_j\big{)}^2 \mu(\ud\vb, t) - \frac{1}{2}\intethree (v_i^2 + v_j^2) \mu(\ud\vb, t)$, define $\ub_{ij} := \frac{\sqrt{2}}{2}\mathbf{e}_i + \frac{\sqrt{2}}{2}\mathbf{e}_j$, and invoke once again (\ref{eq:secondmomSu})-(\ref{eq:secondmomSu1}) to obtain
$$
\intethree v_iv_j \mu(\ud\vb, t) = \et[(S(\ub_{ij}))^2] - \frac{1}{2}\left\{\et[(S(\mathbf{e}_i))^2] + \et[(S(\mathbf{e}_j))^2]\right\} = \epsilon_{ij}(t) + \overline{V}_i\overline{V}_j
$$
for all $t \geq 0$, where $\epsilon_{ij}(t) := e^{-(1 - f_1(b)) t} \intethree (v_i - \overline{V}_i)(v_j - \overline{V}_j) \mu_0(\ud\vb)$, which completes the proof of (\ref{eq:ViVj}).

To prove (\ref{eq:doob}), consider Lemma \ref{lm:petrov} with the same $\gamma$, $G$ and $G_{\ast}$ and define
$$
q(x) := \left\{\begin{array}{ll}
\frac{1}{x^3}\int_{0}^{x}G_{\ast}(y) \ud y & \text{if} \ x > 0 \\
\frac{1}{3} & \text{if} \ x = 0
\end{array} \right.
$$
and $F_{\ast}(x) := x^2q(x)$. This $q$ meets the requirements of the theorem since $\lim_{x \rightarrow +\infty} q(x) = +\infty$ holds after a straightforward application of l'H\^{o}pital's rule, while property ii') of $G$ shows that $q$ is non-decreasing. Then, after noting that $F_{\ast}(x) \leq G_{\ast}(x)$ for all $x \in [0, +\infty)$ thanks to the fact that $G_{\ast}$ is non-decreasing, the combination of Lemma \ref{lm:petrov} with property iv') of $G$ gives
\begin{eqnarray}
\intethree F_{\ast}(|\vb|) \mu(\ud\vb, t) &\leq& \intethree G(|\vb|^2) \mu(\ud\vb, t) = \et[G(S(\mathbf{e}_1)^2 + S(\mathbf{e}_2)^2 + S(\mathbf{e}_3)^2)] \nonumber \\
&\leq& 3\lambda_3^2 \sup_{\ub \in S^2} \et[G_{\ast}(S(\ub))] \leq 3\lambda_3^2 C_1(\mu_0) := C(\mu_0) \nonumber
\end{eqnarray}
which is the desired conclusion. Finally, since (\ref{eq:doob}) is in force, then
$$
\int_{|\vb| \geq R} |\vb|^2 \mu(\ud \vb, t) \leq \frac{1}{q(R)}\intethree |\vb|^2 q(|\vb|) \mu(\ud\vb, t) \leq \frac{C(\mu_0)}{q(R)}
$$
holds for every $t \geq 0$, and the validity of (\ref{eq:unifintcutoff}) follows.

\subsection{Proof of Theorem \ref{thm:main}}

According to the Arkeryd approach, the proof of the existence starts from two inequalities, which are valid for solutions to the SHBEMM with Grad angular cutoff. In fact, after noting that $[b(x) \wedge n]/B_n$ meets (\ref{eq:cutoff}) for all $n \geq n_0 := \min\{n \in \mathbb{N}\ |\ B_n > 0\}$, the Cauchy problem relative to (\ref{eq:boltzmann}), with $[b(x) \wedge n]/B_n$ and $\mu_0$ as collision kernel and initial datum respectively, admits a unique solution $\{\mu_n(\cdot, t)\}_{t \geq 0}$, which possesses all the properties listed in Theorem \ref{thm:cutoff}. In particular, (\ref{eq:Vu0}) yields
\begin{equation} \label{eq:Vu0n}
\intethree \vb \mu_n(\ud\vb, t) = \intethree \vb \mu_0(\ud\vb) := \overline{\mathbf{V}} \ \ \ \text{and} \ \ \ \intethree |\vb|^2 \mu_n(\ud\vb, t) = \mathfrak{m}_2
\end{equation}
for all $t \geq 0$ and $n \geq n_0$, so that the former preliminary inequality reads
\begin{eqnarray}
|\hat{\mu}_n(\xib_2, t) - \hat{\mu}_n(\xib_1, t)| &\leq& |\xib_2 - \xib_1| \sup_{\xib \in \rthree}\big{|}\nabla_{\xib} \hat{\mu}_n(\xib, t)\big{|} \leq |\xib_2 - \xib_1| \intethree |\vb| \mu_n(\ud\vb, t) \nonumber \\
&\leq& |\xib_2 - \xib_1| \Big{(}\intethree |\vb|^2 \mu_n(\ud\vb, t)\Big{)}^{1/2} = \mathfrak{m}_2^{1/2} |\xib_2 - \xib_1| \label{eq:arkeryd1}
\end{eqnarray}
for all $\xib_1, \xib_2 \in \rthree$, with $t$ and $n$ as above. To get the latter inequality, it is worth borrowing
a result from \cite{morimoto} (precisely, Lemma 2.2), re-stated here in a slightly different form.
\begin{lm} \label{lm:morimoto}
\emph{Let} $\chi$ \emph{be a Borel p.m. on} $\rthree$ \emph{such that} $\intethree |\vb|^2 \chi(\ud\vb) < +\infty$ \emph{and}
$b$ \emph{a the collision kernel satisfying} (\ref{eq:cutoff}). \emph{Then},
\begin{equation} \label{eq:morimoto}
\Big{|}\int_{S^2} [\hat{\chi}(\xib_{+}) \hat{\chi}(\xib_{-}) - \hat{\chi}(\xib)] b\left(\frac{\xib}{|\xib|} \cdot \omb \right) \unifSo \Big{|} \leq \frac{3}{2} \overline{B}\ |\xib|^2 \intethree |\vb|^2 \chi(\ud\vb)
\end{equation}
\emph{holds true for all} $\xib \neq \mathbf{0}$, \emph{with} $\xib_{+} := \xib - (\xib \cdot \omb)\omb$, $\xib_{-} := (\xib \cdot \omb)\omb$ \emph{and} $\overline{B} := \int_{0}^{1} x^2 b(x) \ud x$.
\end{lm}
The original formulation in \cite{morimoto} deals with collision kernels satisfying (\ref{eq:vwcutoff}), but, in that case, (\ref{eq:morimoto}) turns out to be false if $\int_{S^2}$ is intended as a standard Lebesgue integral. Indeed, it is enough to choose $\chi$ as a Gaussian probability law (for example, with zero means and covariance matrix $V = (v_{i,j})_{1 \leq i,j \leq 3}$, with $v_{2,2} = v_{3,3} = 1$, $v_{2,3} = v_{3,2} = 1/2$, $v_{i,j} = 0$ otherwise) and $b(x) = |x|^{-5/2}$, to verify that $\int_{S^2} |\hat{\chi}(\xib_{+}) \hat{\chi}(\xib_{-}) - \hat{\chi}(\xib)| b\left(\frac{\xib}{|\xib|} \cdot \omb \right) \unifSo = +\infty$. This counterexample can be easily reformulated also in the different parametrization used in \cite{morimoto}. Therefore, due to its relevance, the original proof of this lemma is shortly reproduced below.

\emph{Proof of Lemma \ref{lm:morimoto}}: Define $\boldsymbol{\zeta} := \left(\xib_{+} \cdot \frac{\xib}{|\xib|}\right) \frac{\xib}{|\xib|}$ and $\tilde{\xib}_{+} := 2\boldsymbol{\zeta} - \xib_{+}$ to write
\begin{eqnarray}
&& \int_{S^2} [\hat{\chi}(\xib_{+}) \hat{\chi}(\xib_{-}) - \hat{\chi}(\xib)] b\left(\frac{\xib}{|\xib|} \cdot \omb \right) \unifSo = \frac{1}{2} \int_{S^2} [\hat{\chi}(\xib_{+}) + \hat{\chi}(\tilde{\xib}_{+}) - 2\hat{\chi}(\boldsymbol{\zeta})] \times \nonumber \\
&\times& b\left(\frac{\xib}{|\xib|} \cdot \omb \right) \unifSo + \int_{S^2} [\hat{\chi}(\boldsymbol{\zeta}) - \hat{\chi}(\xib)] b\left(\frac{\xib}{|\xib|} \cdot \omb \right) \unifSo \nonumber \\
&+& \int_{S^2} \hat{\chi}(\xib_{+})[\hat{\chi}(\xib_{-}) - 1] b\left(\frac{\xib}{|\xib|} \cdot \omb \right) \unifSo \nonumber
\end{eqnarray}
for all $\xib \neq \mathbf{0}$. Upon assuming that $\intethree \vb \chi(\ud\vb) = \mathbf{0}$ -- which does not affect the generality, for the replacement of $\hat{\chi}(\xib)$ with $\hat{\chi}(\xib) \exp\{-i \xib \cdot \intethree \vb \chi(\ud\vb)\}$ does not change the LHS of (\ref{eq:morimoto}) -- invoke the elementary inequality $|\hat{\chi}(\xib) - 1| \leq \frac{1}{2} |\xib|^2 \intethree |\vb|^2 \chi(\ud\vb)$ to obtain
\begin{eqnarray}
\frac{1}{2} |\hat{\chi}(\xib_{+}) + \hat{\chi}(\tilde{\xib}_{+}) - 2\hat{\chi}(\boldsymbol{\zeta})| &\leq& \frac{1}{2} |\xib_{+} - \boldsymbol{\zeta}|^2 \intethree |\vb|^2 \chi(\ud\vb) \nonumber \\
|\hat{\chi}(\boldsymbol{\zeta}) - \hat{\chi}(\xib)| &\leq& \frac{1}{2} |\boldsymbol{\zeta} - \xib|^2 \intethree |\vb|^2 \chi(\ud\vb) \nonumber \\
|\hat{\chi}(\xib_{-}) - 1| &\leq& \frac{1}{2} (\xib \cdot \omb)^2 \intethree |\vb|^2 \chi(\ud\vb)\ , \nonumber
\end{eqnarray}
so that the conclusion is reached by noting that $|\xib_{+} - \boldsymbol{\zeta}| \leq |\xib \cdot \omb|$, $|\boldsymbol{\zeta} - \xib| \leq |\xib \cdot \omb|$
and $\int_{S^2} \left(\frac{\xib}{|\xib|} \cdot \omb \right)^2 b\left(\frac{\xib}{|\xib|} \cdot \omb \right) \unifSo  = \overline{B}$.\ $\square$ \\

At this stage, the Bobylev identity (\ref{eq:bobylev}) shows that
$$
\frac{\partial}{\partial t} \hat{\mu}_n(\xib, t) = \int_{S^2} [\hat{\mu}_n(\xib_{+}, t) \hat{\mu}_n(\xib_{-}, t) - \hat{\mu}_n(\xib, t)] \frac{b(\xib/|\xib| \cdot \omb) \wedge n}{B_n}\unifSo
$$
is valid for all $\xib \neq \mathbf{0}$, $t > 0$ and $n \geq n_0$, which, in combination with Lemma \ref{lm:morimoto} and (\ref{eq:Vu0n}), yields
\begin{equation}\label{eq:arkeryd2}
|\hat{\mu}_n(\xib, B_n t_2) - \hat{\mu}_n(\xib, B_n t_1)| = \Big{|}\int_{B_n t_1}^{B_n t_2}\Big{[}
\frac{\partial}{\partial s} \hat{\mu}_n(\xib, s)\Big{]} \ud s \Big{|} \leq \frac{3}{2} \overline{B}\ \mathfrak{m}_2 |\xib|^2 |t_2 - t_1|
\end{equation}
for all $\xib \in \rthree$, $t_1, t_2 \geq 0$ and $n \geq n_0$, corresponding to the latter preliminary inequality of the Arkeryd approach.

To proceed, select a sequence $T := \{t_k\}_{k \geq 1}$, dense in $[0,+\infty)$, and note that $\{\hat{\mu}_n(\xib, B_n t_k)\}_{n \geq n_0, k \in \mathbb{N}}$ is a \emph{uniformly bounded} and \emph{equicontinuous} family of complex-valued functions on $\rthree$, since the former property follows from $|\hat{\mu}_n(\xib, B_n t_k)| \leq 1$, while the latter is a consequence of (\ref{eq:arkeryd1}). Hence, the Ascoli-Arzel\`a theorem, combined with the L\'{e}vy continuity theorem and the Cantor diagonal argument, ensures the existence of a sequence $\{\mu(\cdot, t_k)\}_{k \geq 1}$ of Borel p.m.'s on $\rthree$ such that, for all $t_k \in T$, $\mu_{n_l}(\cdot, B_{n_l} t_k) \Rightarrow \mu(\cdot, t_k)$ as $l \rightarrow +\infty$, $\{n_l\}_{l \geq 1}$ being a suitable subsequence of integers which diverges at infinity. For any other $t \in [0,+\infty)\setminus T$, take any subsequence $\{t_{k_r}\}_{r \geq 1} \subset T$, converging to $t$, and consider $\text{w-lim}_{r \rightarrow +\infty} \mu(\cdot, t_{k_r})$: This limit exists and is independent of the choice of the approximating sequence $\{t_{k_r}\}_{r \geq 1}$, since (\ref{eq:arkeryd2}) entails $|\hat{\mu}(\xib, t^{''}) - \hat{\mu}(\xib, t^{'})| \leq \frac{3}{2}\overline{B}\ \mathfrak{m}_2 |\xib|^2 |t^{''} - t^{'}|$ for all $\xib \in \rthree$ and $t^{'}, t^{''} \in T$. More precisely, $\{\hat{\mu}(\xib, t_{k_r})\}_{r \geq 1}$ is a Cauchy sequence in $\mathbb{C}$ which converges to some $g_t(\xib)$, for any fixed $\xib \in \rthree$, the limit being continuous as a function of $\xib$ thanks to the inequality $|g_t(\xib_2) - g_t(\xib_1)| \leq \mathfrak{m}_2^{1/2} |\xib_2 - \xib_1|$, which obviously emanates from the combination of (\ref{eq:arkeryd1}) with (\ref{eq:arkeryd2}). A further application of the L\'{e}vy continuity theorem shows that there exists, for all $t \in [0,+\infty)\setminus T$, a Borel p.m. $\mu(\cdot, t)$ on $\rthree$ such that $g_t(\xib) = \hat{\mu}(\xib, t)$ holds for all $\xib \in \rthree$ and that $\mu(\cdot, t_{k_r}) \Rightarrow \mu(\cdot, t)$, as $r \rightarrow +\infty$. The conclusion of this line of reasoning is that $\mut$ satisfies
\begin{enumerate}
\item[A)] $\mu_{n_l}(\cdot, B_{n_l} t) \Rightarrow \mu(\cdot, t)$ as $l \rightarrow +\infty$, for all $t \geq 0$;
\item[B)] $|\hat{\mu}(\xib, t^{''}) - \hat{\mu}(\xib, t^{'})| \leq \frac{3}{2}\overline{B}\ \mathfrak{m}_2 |\xib|^2 |t^{''} - t^{'}| $ for all $\xib \in \rthree$ and $t^{'}, t^{''} \in [0,+\infty)$,
\end{enumerate}
$\{\mut\}_{t \geq 0}$ being the obvious candidate as solution of (\ref{eq:boltzmann}). Indeed, $\mu(\cdot, 0) = \mu_0(\cdot)$ holds by A), while $\intethree |\vb|^2 \mu(\ud\vb, t) < +\infty$ and (\ref{eq:Vu0}) are in force, for all $t \geq 0$, as a consequence of Lemma 1 in \cite{tovil}, whose hypotheses are fulfilled in view of A) and
\begin{equation} \label{eq:doobnl}
\sup_{\substack{l \in \mathbb{N} \\ t \geq 0}} \ \int_{|\vb| \geq R} |\vb|^2 \mu_{n_l}(\ud\vb, B_{n_l} t) \leq \frac{1}{q(R)} \sup_{\substack{l \in \mathbb{N} \\ t \geq 0}} \ \intethree |\vb|^2 q(|\vb|) \mu_{n_l}(\ud\vb, B_{n_l} t) \leq \frac{C(\mu_0)}{q(R)}\ ,
\end{equation}
which emanates from (\ref{eq:doob}). After fixing $\psi \in \DD$, Definition \ref{defin:solution} entails
\begin{eqnarray}
&& \intethree \psi(\vb) \mu_{n_l}(\ud\vb, B_{n_l} t) = \intethree \psi(\vb) \mu_0(\ud\vb) \nonumber \\
&+& \int_{0}^{t} \intethree\intethree \int_{-1}^{1} A_{\psi}(\vb, \wb, \xi) \mu_{n_l}(\ud\vb, B_{n_l} \tau) \mu_{n_l}(\ud\wb, B_{n_l} \tau) \xi^2 [b(\xi) \wedge n_l]\ud\xi \ud \tau \ \ \ \ \ \label{eq:boltzmannintegraln}
\end{eqnarray}
for all $t \geq 0$ and $l \in \mathbb{N}$, where
\begin{eqnarray}
A_{\psi}(\vb, \wb, \xi) &:=& \frac{1}{8\pi} \int_{0}^{2\pi} \int_{0}^{1}  \ind\{\vb \neq \wb\} (1 - s) \Big{[} \nabla \psi(\vb_{\ast}(s \xi)) \cdot \frac{\ud^2 \vb_{\ast}}{\ud x^2}(s \xi) \nonumber \\
&+& \nabla \psi(\wb_{\ast}(s \xi)) \cdot \frac{\ud^2 \wb_{\ast}}{\ud x^2}(s \xi) + \left(\frac{\ud \vb_{\ast}}{\ud x}(s \xi)\right)^t \mathrm{Hess}[\psi](\vb_{\ast}(s \xi))\left(\frac{\ud \vb_{\ast}}{\ud x}(s \xi)\right)  \nonumber \\
&+& \left(\frac{\ud \wb_{\ast}}{\ud x}(s \xi)\right)^t \mathrm{Hess}[\psi](\wb_{\ast}(s \xi))\left(\frac{\ud \wb_{\ast}}{\ud x}(s \xi)\right)\Big{]} \ud s\ud\theta\ .  \nonumber
\end{eqnarray}
The bounds provided in Subsection \ref{sect:villani} give
\begin{equation}\label{eq:boundApsi}
|A_{\psi}(\vb, \wb, \xi)| \leq K_{\psi}(1 + |\vb - \wb|^2)
\end{equation}
for all $(\vb, \wb, \xi) \in \rone^6 \times [-1,1]$ with a suitable positive constant $K_{\psi}$, while the same argument contained in Subsection \ref{sect:proofextendedQ} shows that $(\vb, \wb) \mapsto \int_{-1}^{1} A_{\psi}(\vb, \wb, \xi)\\ \xi^2 b(\xi)\ud\xi$ is continuous on $\rone^6$. The key point consists now in exploiting A) to take the limit of both sides of (\ref{eq:boltzmannintegraln}) as $n_l \rightarrow +\infty$, with particular attention to the multiple integral on the RHS, which
will be proved to converge to
$$
\int_{0}^{t} \intethree\intethree \int_{-1}^{1} A_{\psi}(\vb, \wb, \xi) \mu(\ud\vb, \tau) \mu(\ud\wb, \tau) \xi^2 b(\xi)\ud\xi \ud \tau\ .
$$
Indeed, thanks to the dominated convergence theorem, combined with (\ref{eq:vwcutoff}), (\ref{eq:Vu0n}) and (\ref{eq:boundApsi}), it is enough to show that both the quantities
$$
\Big{|}\intethree\intethree \int_{-1}^{1} A_{\psi}(\vb, \wb, \xi) \mu_{n_l}(\ud\vb, B_{n_l} \tau) \mu_{n_l}(\ud\wb, B_{n_l} \tau) \xi^2 [b(\xi) - (b(\xi) \wedge n_l)]\ud\xi \Big{|}
$$
and
$$
\Big{|}\intethree\intethree \Big{(}\int_{-1}^{1} A_{\psi}(\vb, \wb, \xi) \xi^2 b(\xi)\ud\xi \Big{)} [\mu_{n_l}(\ud\vb, B_{n_l} \tau) \mu_{n_l}(\ud\wb, B_{n_l} \tau) - \mu(\ud\vb, \tau) \mu(\ud\wb, \tau)] \Big{|}
$$
go to zero for all $\tau \geq 0$, as $n_l \rightarrow +\infty$. Apropos of the former, use (\ref{eq:Vu0n}) and (\ref{eq:boundApsi}) to bound it from above by $K_{\psi}(1 + 4\mathfrak{m}_2) \int_{-1}^{1} \xi^2 [b(\xi) - (b(\xi) \wedge n_l)]\ud\xi$, which goes to zero by (\ref{eq:vwcutoff}). As to the latter, note in advance that $\mu_{n_l}(\cdot, B_{n_l} \tau) \otimes \mu_{n_l}(\cdot, B_{n_l} \tau) \Rightarrow \mu(\cdot, \tau) \otimes \mu(\cdot, \tau)$ thanks to Theorem 2.8 in \cite{bill2},
and that
$$
\lim_{R \rightarrow +\infty} \sup_{l \in \mathbb{N}} \int_{|\vb|^2 + |\wb|^2 \geq R} (|\vb|^2 + |\wb|^2) \mu_{n_l}(\ud\vb, B_{n_l} \tau) \mu_{n_l}(\ud\wb, B_{n_l} \tau) = 0
$$
in view of Lemma 1 in \cite{tovil}. Thus, an application of Theorem 7.12 in \cite{vilMass} leads to the desired conclusion about the asymptotic behavior of the latter quantity at issue. Whence,
\begin{eqnarray}
\intethree \psi(\vb) \mu(\ud\vb, t) &=& \intethree \psi(\vb) \mu_0(\ud\vb) \nonumber \\
&+& \int_{0}^{t} \intethree\intethree \int_{-1}^{1} A_{\psi}(\vb, \wb, \xi) \mu(\ud\vb, \tau) \mu(\ud\wb, \tau) \xi^2 b(\xi) \ud\xi \ud \tau \label{eq:boltzmannintegral}
\end{eqnarray}
holds for all $t \geq 0$, by which $t \mapsto \intethree \psi(\vb) \mu(\ud\vb, t)$ turns out to be continuous on $[0, +\infty)$. Lastly, take a sequence $\{\tau_k\}_{k \geq 1} \subset [0, +\infty)$, converging to some given $\tau \in [0, +\infty)$, and mimic the above argument to obtain that $\mu(\cdot, \tau_k) \otimes \mu(\cdot, \tau_k) \Rightarrow \mu(\cdot, \tau) \otimes \mu(\cdot, \tau)$ and that
\begin{gather}
\lim_{k \rightarrow +\infty} \intethree\intethree \Big{(}\int_{-1}^{1} A_{\psi}(\vb, \wb, \xi) \xi^2 b(\xi)\ud\xi \Big{)} \mu(\ud\vb, \tau_k) \mu(\ud\wb, \tau_k) \nonumber \\
= \intethree\intethree \Big{(}\int_{-1}^{1} A_{\psi}(\vb, \wb, \xi) \xi^2 b(\xi)\ud\xi \Big{)}\mu(\ud\vb, \tau) \mu(\ud\wb, \tau) \ . \nonumber
\end{gather}
Combining this continuity with (\ref{eq:boltzmannintegral}) proves that $t \mapsto \intethree \psi(\vb) \mu(\ud\vb, t)$ is also continuously differentiable on $(0, +\infty)$ and that (\ref{eq:wboltzmann})-(\ref{eq:vwRHS}) are in force for any fixed $\psi \in \DD$.

To conclude the proof, consider the additional properties of the solution $\mut$, just obtained as limit of $\{\mu_{n_l}(\cdot, B_{n_l} t)\}_{l \geq 1}$. First, the identities proved in Subsection \ref{sect:moments} lead to
\begin{eqnarray}
\intethree v_i^2 \mu_{n_l}(\ud\vb, B_{n_l} t) - \overline{V}_i^2 &=&  e_{n_l}(t) \Big{[} \intethree \{(v_i - \overline{V_i})^2 - \frac{1}{3}|\vb - \overline{\mathbf{V}}|^2\} \mu_0(\ud\vb)\Big{]} \nonumber \\
&+& \frac{1}{3} \intethree |\vb - \overline{\mathbf{V}}|^2 \mu_0(\ud\vb) \label{eq:wchaosn1} \\
\intethree v_i v_j \mu_{n_l}(\ud\vb, B_{n_l} t) - \overline{V}_i \overline{V}_j &=& e_{n_l}(t)\intethree (v_i - \overline{V}_i)(v_j - \overline{V}_j) \mu_0(\ud\vb) \label{eq:wchaosn2}
\end{eqnarray}
where
$$
e_{n_l}(t) = \exp\Big{\{}-\frac{3}{2}\Big{[} 2\int_{0}^{1} x^2(1-x^2)\frac{b(x) \wedge n_l}{B_{n_l}} \ud x \Big{]} B_{n_l}t\Big{\}} \ .
$$
The uniform integrability of the second absolute moments of the $\mu_{n_l}$'s, encapsulated in (\ref{eq:doobnl}), yields
$\lim_{l \rightarrow +\infty} \intethree v_i v_j \mu_{n_l}(\ud\vb, B_{n_l} t) = \intethree v_i v_j \mu(\ud\vb, t)$ for all $i, j \in \{1, 2, 3\}$ and $t \geq 0$, while an obvious application of the monotone convergence theorem shows that $\lim_{l \rightarrow +\infty} e_{n_l}(t) = \exp\{-\frac{3}{2}[2\int_{0}^{1} x^2(1-x^2)b(x) \ud x] t\}$ for all $t \geq 0$. Hence, (\ref{eq:wchaosn1})-(\ref{eq:wchaosn2}) pass to the limit as $l \rightarrow +\infty$, and (\ref{eq:ViVj}) follows. Apropos of the extension of (\ref{eq:doob}), write
\begin{eqnarray}
\intethree \min\{|\vb|^2 q(|\vb|), m\} \mu(\ud\vb, t) &=& \lim_{l \rightarrow +\infty} \intethree \min\{|\vb|^2 q(|\vb|), m\} \mu_{n_l}(\ud\vb, B_{n_l} t) \nonumber \\
&\leq& \sup_{\substack{l \in \mathbb{N} \\ t \geq 0}} \ \intethree |\vb|^2 q(|\vb|) \mu_{n_l}(\ud\vb, B_{n_l} t) \leq C(\mu_0) \nonumber
\end{eqnarray}
for all $m \in \mathbb{N}$. Thus, the monotone convergence theorem shows that (\ref{eq:doob}) continues to be valid with the same $q$ and $C(\mu_0)$ as in Theorem \ref{thm:cutoff}, and (\ref{eq:unifintcutoff}) follows by the very same argument used at the end of Subsection \ref{sect:moments}. Finally, take $R$ and $f_R$ as in Lemma \ref{lm:extendedQ} to obtain that the only solution of (\ref{eq:boltzmann}) with $\frac{b \wedge n_l}{B_{n_l}}$ and $\mu_0 \circ f_{R}^{-1}$ as collision kernel and initial datum, respectively, is $\{\mu_{n_l}(\cdot, B_{n_l} t) \circ f_{R}^{-1}\}_{t \geq 0}$, in view of Theorem \ref{thm:cutoff}. Since
$\mu_{n_l}(\cdot, B_{n_l} t) \circ f_{R}^{-1} \Rightarrow \mut \circ f_{R}^{-1}$ is valid for all $t \geq 0$, as $l \rightarrow +\infty$, in view of the continuous mapping theorem (cf. Theorem 2.7 in \cite{bill2}), then $\mut \circ f_{R}^{-1}$ is a solution of (\ref{eq:boltzmann}) with $b$ and $\mu_0 \circ f_{R}^{-1}$ as collision kernel and initial datum, respectively.


\begin{thebibliography}{99}

\bibitem{abra} Abrahamsson, F.: Strong $L^1$ convergence to equilibrium without entropy conditions for the Boltzmann equation. Comm. Partial Differential Equations $\textbf{24}$, 1501-1535 (1999)

\bibitem{ark} Arkeryd, L.: Intermolecular forces of infinite range and the Boltzmann equation. Arch. Rational Mech. Anal.  $\textbf{77}$, 11-21 (1981)

\bibitem{bill2} Billingsley, P.: Convergence of Probability Measures. $2^{nd}$ ed. Wiley, New York (1999)

\bibitem{bob88} Bobylev, A.V.: The theory of the nonlinear spatially uniform Boltzmann equation for Maxwell molecules. Mathematical Physics Reviews $\textbf{7}$, 111-233 (1988)

\bibitem{caka} Cannone, M., Karch, G.: Infinite energy solutions to the homogeneous Boltzmann equation. Comm. Pure Appl. Math. $\textbf{63}$, 747-778 (2010)

\bibitem{cl} Carlen, E.A., Lu, X.: Fast and slow convergence to equilibrium for Maxwellian molecules via Wild sums. J. Statist. Phys. $\mathbf{112}$, 59-134 (2003)

\bibitem{cerS} Cercignani, C.: The Boltzmann Equation and its Applications. Springer-Verlag, New York (1988)

\bibitem{cip} Cercignani, C., Illner, R., Pulvirenti, M.: The Mathematical Theory of Dilute Gases. Springer-Verlag, New York (1994)

\bibitem{desvKac} Desvillettes, L.: About the regularizing properties of the non-cut-off Kac equation. Comm. Math. Phys. $\textbf{168}$, 417-440 (1995)

\bibitem{doob} Doob, J.L.: Measure Theory. Springer Verlag, New York (1994)

\bibitem{doreACTA} Dolera, E., Regazzini, E.: Proof of a McKean conjecture on the rate of convergence of Boltzmann-equation solutions. Submitted. arXiv:1206.5147 (2012)

\bibitem{fou} Fournier, N.: Existence and regularity study for two-dimensional Kac equation without cutoff by a probabilistic approach. Ann. Appl. Probab. $\textbf{2}$, 434-462 (2000)

\bibitem{gtw} Gabetta, E., Toscani, G., Wennberg, B.: Metrics for probability distributions and the trend to equilibrium for solutions of the Boltzmann equation. J. Statist. Phys. $\textbf{81}$, 901-934 (1995)

\bibitem{goudon} Goudon, T.: On Boltzmann equations and Fokker-Planck asymptotics: influence of grazing collisions. J. Statist. Phys. $\textbf{89}$, 751-776 (1997)

\bibitem{iktr} Ikenberry, E., Truesdell, C.: On the pressures and the flux of energy in a gas according to Maxwell's kinetic theory.I. J. Rational Mech. Anal. $\textbf{5}$, 1-54 (1956)

\bibitem{max} Maxwell, J.C.: On the dynamical theory of gases. Philos. Trans. Roy. Soc. London Ser. A \textbf{157}, 49-88 (1867)

\bibitem{mck7} McKean, H.P.Jr.: An exponential formula for solving Boltzmann's equation for a Maxwellian gas. J. Combinatorial Theory $\textbf{2}$, 358-382 (1967)

\bibitem{meyer} Meyer, C.: Probability and Potential. Springer-Verlag, New York (1988)

\bibitem{miwe} Mischler, S., Wennberg, B.: On the spatially homogeneous Boltzmann equation. Ann. Inst. H. Poincar\'{e} Anal. Non Lin\'{e}aire $\textbf{16}$, 467-501 (1999).

\bibitem{mor} Morgenstern, D.: General existence and uniqueness proof for the spatially homogeneous solutions of the Maxwell-Boltzmann equation in the case of Maxwellian molecules. Proc. Nat. Acad. Sci. USA \textbf{40}, 719-721 (1954)

\bibitem{morimoto} Morimoto, Y.: A remark on Cannone-Karch solutions to the homogeneous Boltzmann equation for Maxwellian molecules. Kinet. Relat. Models $\textbf{5}$, 551-561 (2012)

\bibitem{pe1} Petrov, V.V.: Limit Theorems of Probability Theory. Sequences of Independent Random Variables. The Clarendon Press, Oxford University Press, New York (1995)

\bibitem{puto} Pulvirenti, A., Toscani, G.: The theory of the nonlinear Boltzmann equation for Mawxell molecules in Fourier representation. Ann. Mat. Pura Appl. $\textbf{171}$, 181-204 (1996)

\bibitem{ta} Tanaka, H.: Probabilistic treatement of the Boltzmann equation of Maxwellian molecules. Z. Wahrsch. Verw. Gebiete $\textbf{46}$, 67-105 (1978)

\bibitem{tovil} Toscani, G., Villani, C.: Probability metrics and uniqueness of the solution of the Boltzmann equation for a Maxwell gas. J. Statist. Phys. $\textbf{94}$, 619-637 (1999)

\bibitem{villNew} Villani, C.: On a new class of weak solutions to the spatially homogeneous Boltzmann and Landau equations. Arch. Rational Mech. Anal. $\textbf{143}$, 273-307 (1998)

\bibitem{vil} Villani, C.: A review of mathematical topics in collisional kinetic theory. In: Handbook of Mathematical Fluid Dynamics, vol. 1, pp. 71-305. (S. Friedlander and D. Serre eds.), North-Holland, Amsterdam (2002)

\bibitem{vilMass} Villani, C.: Topics in optimal transportation. American Mathematical Society, Providence (2003)

\bibitem{wil} Wild, E.: On Boltzmann's equation in kinetic theory of gases. Proc. Cambridge Philos. Soc. \textbf{47}, 602-609 (1951)
\end{thebibliography}
\end{document}